\documentclass{article}
\pdfoutput=1
\usepackage{graphicx}

\usepackage{jheppub}
\usepackage{amsmath}
\usepackage{epsfig,multicol,bbm}
\usepackage{url}

\usepackage{graphicx}
\usepackage{setspace}
\usepackage{amssymb}
\usepackage{listings}
\usepackage{epstopdf}
\usepackage{float}
\usepackage{color}
\usepackage{amsmath}
\usepackage{ dsfont }
\usepackage{slashed}
\usepackage{subfigure}

\newcommand{\newc}{\newcommand}
\newc{\gsim}{\lower.7ex\hbox{$\;\stackrel{\textstyle>}{\sim}\;$}}
\newc{\lsim}{\lower.7ex\hbox{$\;\stackrel{\textstyle<}{\sim}\;$}}
\newc{\gev}{\,{\rm GeV}}
\newc{\mev}{\,{\rm MeV}}
\newc{\ev}{\,{\rm eV}}
\newc{\kev}{\,{\rm keV}}
\newc{\tev}{\,{\rm TeV}}

\def\ln{\mathop{\rm ln}}

\newc{\mz}{M_Z}
\newc{\mpl}{M_*}
\newc{\mw}{m_{\rm weak}}
\newc{\nr}[1]{N^c_R{}_{#1}}
\usepackage{amsmath}

\def\bitem{\begin{itemize}}
\def\eitem{\end{itemize}}
\newc{\ie}{{\it i.e.}}          \newc{\etal}{{\it et al.}}
\newc{\eg}{{\it e.g.}}          \newc{\etc}{{\it etc.}}
\newc{\cf}{{\it c.f.}}

\newcommand{\CO}{\mathcal{O}}

\newcommand\fverb{\setbox\fverbbox=\hbox\bgroup\verb}
\newcommand\fverbdo{\egroup\medskip\noindent%
            \fbox{\unhbox\fverbbox}\ }
\newcommand\fverbit{\egroup\item[\fbox{\unhbox\fverbbox}]}
\newbox\fverbbox

\numberwithin{equation}{section}

\long\def\symbolfootnote[#1]#2{\begingroup%
\def\thefootnote{\fnsymbol{footnote}}\footnote[#1]{#2}\endgroup}

\newcommand{\be}{\begin{equation}}
\newcommand{\ee}{\end{equation}}

\newcommand{\bea}{\begin{eqnarray}\begin{aligned}}
\newcommand{\eea}{\end{aligned}\end{eqnarray}}
\newcommand{\mat}{\begin{pmatrix}}
\newcommand{\rix}{\end{pmatrix}}

\renewcommand{\bar}{\overline}
\renewcommand{\slash}[1]{#1\!\!\!/}

\newcommand{\go}{{\tilde g}}
\newcommand{\Bo}{{\tilde B}}

\newcommand{\Ho}{{\tilde H}}

\newcommand{\cho}{{\tilde \chi}}

\newcommand{\st}{{\tilde t}}
\newcommand{\sbo}{{\tilde b}}

\newcommand{\stau}{{\tilde\tau}}

\newcommand{\qq}{\qquad}

\renewcommand{\d}{\partial}

\newcommand{\beqa}{\begin{eqnarray}}
\newcommand{\eeqa}{\end{eqnarray}}
\newcommand{\lp}{\left(}
\newcommand{\rp}{\right)}

\newcommand{\beq}{\begin{equation}}
\newcommand{\eeq}{\end{equation}}
\newcommand{\hc}{\mbox{h.c.}}
\newcommand{\braced}[1]{$\{#1\}$}

\newcommand{\abs}[1]{\left\vert#1\right\vert}

\newcommand{\order}[1]{{\cal O}\left(#1\right)}

\newcommand{\met}{{\slash E_T}}

\newcommand{\LoM}{\frac \Lambda M}

\newcommand{\spacer}{\LARGE {\color{white}I}\hspace{-2mm}\normalsize}

\newcommand{\CQ}{\mathcal{Q}}

\newcommand{\CL}{\mathcal{L}}

\title{ Surveying Extended GMSB Models with $m_h=125$ GeV}

\date{\today}

\author[]{Jared A.\ Evans and}
\author[]{David Shih}

\affiliation[]{NHETC \\ Department of Physics and Astronomy\\
Rutgers University \\
Piscataway, NJ 08854 }

\preprint{RUNHETC-2013-05}    

\abstract{In order to achieve maximal stop mixing and $m_h=125$ GeV, we consider extensions of minimal GMSB that include marginal MSSM-messenger superpotential interactions. Using a new approach to analytic continuation in superspace, we derive general formulas for the soft masses in the presence of such interactions, correctly taking into account the role of MSSM-messenger mixing in a general framework for the first time. We classify and catalog all possible such interactions consistent with perturbative $SU(5)$ unification, and we survey the impact of turning on one interaction at a time, from the point of view of fine tuning, spectrum and phenomenology. We find that the best models are fine-tuned to the sub-percent level and are accessible at the 14 TeV LHC. We highlight potential search strategies that can probe the characteristic spectra of these models. }

\begin{document}

\maketitle

\section{Introduction}
\label{sec:intro}

Recently, both the ATLAS and CMS experiments have announced the discovery of a Standard Model-like Higgs with $m_h\approx 125$ GeV~\cite{Aad:2012tfa,Chatrchyan:2012ufa}. This exciting result provides interesting hints and challenges for physics beyond the Standard Model. Supersymmetry (SUSY), long the preferred solution to the hierarchy problem, is highly constrained by this value of the Higgs mass. In particular, in the MSSM, large radiative corrections from stop/top loops  are needed for $m_h=125$ GeV (see e.g.\ \cite{Hall:2011aa, Heinemeyer:2011aa,Arbey:2011ab,Draper:2011aa,Carena:2011aa}).  These contributions can arise either through extremely heavy, unmixed stops ($M_{S}\equiv\sqrt{m_{\tilde t_1}m_{\tilde t_2}}\gtrsim 10$ TeV), or through lighter stops with maximal mixing~\cite{Casas:1994us,Carena:1995bx,Haber:1996fp}:
\begin{equation}
A_t\sim \sqrt{6}\,M_{S}\quad {\rm and}\quad M_{S}\gtrsim 1\,\,\,{\rm TeV}
\end{equation}
By transplanting the stops from 1 to 10 TeV, the theory grows two orders of magnitude more tuned.  Since such a tuned model has little hope for ever being observed at the LHC, we will focus on generating the light, mixed stops in this work.

Large $A$-terms are essential for obtaining a heavy Higgs with lighter stops. This presents a special challenge for models of gauge mediated SUSY breaking (GMSB) (for a review and original references, see \cite{Giudice:1998bp}), which are strongly motivated by the SUSY flavor problem, but do not produce $A$-terms at the messenger scale.  Large $A$-terms can be generated through RG running driven by a heavy gluino \cite{Draper:2011aa}, but this requires a very large messenger scale and, again, reduces both the naturalness and the likelihood of observing any superpartners at the LHC.  (See also \cite{Ajaib:2012vc}, which studies these issues in the context of minimal GMSB and reaches the same conclusions.)

In this paper, we will instead study models which directly generate large $A_t$ at low messenger scales through marginal superpotential interactions between MSSM and messenger superfields.  We will focus on fully calculable models of perturbative messengers coupled to SUSY-breaking spurions. There are then two types of marginal interaction terms: MSSM-messenger-messenger couplings, and MSSM-MSSM-messenger couplings.  We will refer to these couplings as type I and type II couplings, respectively. It is useful to further divide the type I couplings into two distinct subclasses, those in which the MSSM superfield participating in the interaction is a Higgs, and those in which it is a squark.

It is also useful to distinguish between couplings which give rise to mixing between the MSSM and messenger fields and couplings which do not.  A prime example of a model with MSSM-messenger mixing is the $W=\lambda Q_3U_3\Phi$ model, where $\Phi$ is a messenger with the same quantum numbers as $H_u$. Other similar examples include $W=\lambda H_u Q_3  \Phi$ ($\Phi$ mixing with $U_3$) and $W=\lambda H_u  U_3 \Phi$ ($\Phi$ mixing with $Q_3$). Such models have been studied by many authors in the literature, including  \cite{ChackoPonton,Shadmi:2011hs,Evans:2011bea,Jelinski:2011xe,Evans:2012hg,Albaid:2012qk,Abdullah:2012tq,Perez:2012mj,Endo:2012rd}. We only consider mixing in type II interactions, as the messengers in type I interactions can always be charged with a parity symmetry to forbid mixing.  

\begin{figure}[b]
\begin{center}
\includegraphics[width=6.2in]{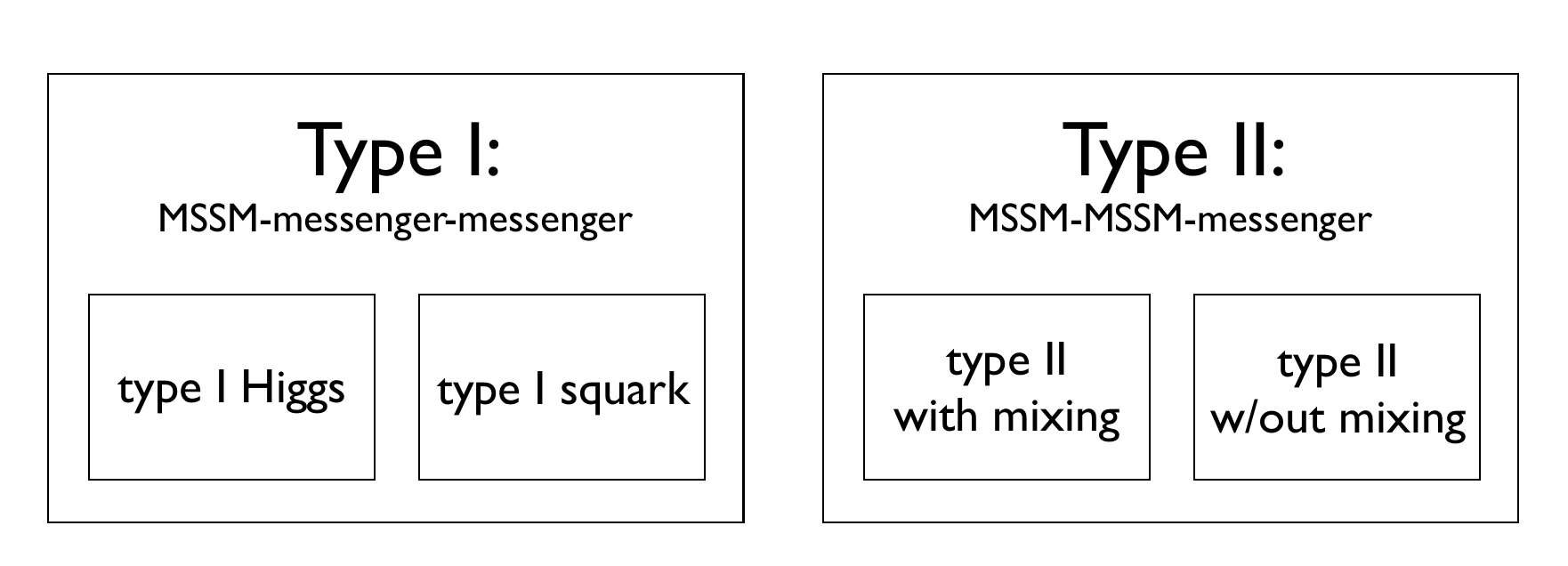}
\caption{A diagram illustrating our classification of models.}
\label{fig:classification}
\end{center}
\end{figure}

Fig.\ \ref{fig:classification} displays this classification of models -- into type I Higgs, type I squark, type II with mixing and type II without mixing. This classification will form the basis for the results presented in this paper. A primary goal of this work will be to describe models within these different categories and their general features with regards to fine tuning and phenomenology.

The main theoretical challenge facing calculable models for large $A$-terms  is something that was dubbed {\it the  $A/m^2$ problem} in \cite{CraigShih} (see also \cite{Craig:2013wga} for a recent discussion in a more general context). In direct analogy to the $\mu/B_\mu$ problem, models for $A$-terms tend to also generate one-loop soft mass-squareds. Such large soft masses would be disastrous for naturalness and/or electroweak symmetry breaking (EWSB). 
As shown in \cite{CraigShih} (following \cite{Giudice:2007ca}), spurion models which avoid the $A/m^2$ problem must be of the minimal gauge mediation (MGM) type:
\begin{equation}
W_{MGM} = \kappa X \Phi\tilde\Phi
\end{equation}
in which all mass-scales in the messenger sector (consisting here of $\Phi$ and $\tilde\Phi$) originate from a single spurion $X$ with $\langle X\rangle = M+ \theta^2 F$. All of the models that we consider in this paper are assumed to have a messenger sector of this form.

Although $A$-terms are commonly viewed as trilinear soft terms, they actually arise as {\it bilinear} terms between the MSSM fields and their $F$-components:
\bea
\label{eq:bilinearA}
\CL \supset A_{H_u} F_{H_u}^\dagger H_u + A_{Q_3} F_{Q_3}^\dagger Q_3 + A_{U_3} F_{U_3}^\dagger U_3  
\eea
After integrating out the $F$-components, this becomes $A_t H_u Q_3 U_3$ with
\bea
\label{eq:Atsum}
A_t = y_t (A_{H_u}+A_{Q_3}+A_{U_3})
\eea
In order to generate $A_t$, it is mandatory that either $H_u$, $Q_3$, or $U_3$ participate in the direct interactions with the messengers. In \cite{CraigShih}, only $H_u$ couplings to messengers were considered, since this automatically preserves minimal flavor violation (MFV), one of the best features of gauge mediation. However, these models suffer from the residual ``little $A/m^2$ problem"  \cite{CraigShih}: by integrating out $F_{H_u}$ in (\ref{eq:bilinearA}), an irreducible contribution $\delta m_{H_u}^2=+A_{H_u}^2$ is generated. Since $A_{H_u}\gtrsim 2$ TeV is required for maximal mixing and $m_h=125$ GeV, this results  in an irreducible fine tuning at the $\sim 10^4$ level in these models.

Motivated by these considerations, we will broaden the scope of \cite{CraigShih} and survey  the complete class of spurion-messenger models for large $A_t$, including also non-MFV couplings involving $Q_3$ and $U_3$. As expected, these squark-type couplings result in far less fine-tuning than the Higgs-type couplings, since they do not generate a little $A/m_{H_u}^2$ problem. For simplicity, and to minimize flavor changing effects, we consider scenarios with only a single coupling introduced between MSSM and messenger fields. This structure may seem unrealistic; however, it is technically natural and facilitates focus on the ``best-case scenarios."  By restricting the models to complete $SU(5)$ messenger multiplets and requiring perturbative gauge couplings up to the GUT scale, only a finite and manageable list of couplings is permitted. Under these assumptions, there are a total of 31 couplings.  The complete list of these can be found in table \ref{table:models}. 

For computing the soft masses in these models, general formulas were derived in \cite{ChackoPonton} using wavefunction renormalization and the technique of ``analytic continuation into superspace" \cite{Giudice:1997ni}. These formulas  are cast  in terms of anomalous dimensions and beta functions of non-holomorphic couplings, and they are sufficient for all type I and most type II models.  However, one runs into numerous complications when attempting to apply them to type II models with mixing between MSSM and messenger superfields (such as the $QU\Phi$ model described above). The problem stems from crucial assumption made in  \cite{ChackoPonton} that the wavefunctions are continuous through the messenger threshold. This proves to be incompatible with  standard conventions for the beta functions and anomalous dimensions (e.g.\ those in \cite{MartinVaughn}). Attempting to substitute standard beta functions and anomalous dimensions into the general formulas of \cite{ChackoPonton}, as many in the literature have done, leads to incorrect results.\footnote{While one can correctly apply the formulas of \cite{ChackoPonton} in the mixed type II case by using non-standard beta functions and anomalous dimensions (see \cite{Evans:2012hg}), this has only been performed in one specific model, and standardized formulas do not exist.}  In fact, we have found these complications to be so insidious that nearly every paper that studies models with MSSM-messenger mixing contains erroneous formulas.

Faced with many incorrect results and an enormous amount of confusion in the literature, we will devote section 2 to describing a new approach to deriving the soft masses using wavefunction renormalization. The main idea of our new approach is to obtain the wavefunctions by direct integration  
in a manifestly holomorphic scheme,  rather than first utilizing the beta functions for the non-holomorphic couplings, as done in \cite{ChackoPonton}. This results in a conceptually cleaner approach, which circumvents the difficulties of \cite{ChackoPonton} and leads to fully general, correct formulas which can be applied uniformly to all models, whether mixing is present or not.  In appendix A, we provide many checks of our results for the $QU\Phi$ model described above, and illustrate the complications that arise in type II models with mixing, using this concrete example.

With the correct and fully general formulas for the soft masses in hand, we will compute the soft masses at the messenger scale for each model in table \ref{table:models}, and then investigate the parameter space where $m_h=125$ GeV.  For such points in the viable parameter space, we will survey the phenomenology and fine tuning. In order to quantify tuning in these models, we utilize a tuning measure, $\Delta_{FT}$, that is based on the Giudice-Barbieri tuning measure \cite{GiudiceBarbieri}, but with a slightly unconventional choice of underlying parameters. This choice is made to ensure that the tuning measure captures all sensitivities successfully, does not introduce artificial tuning, and assigns comparable weight to uncorrelated contributions which cancel against one another.  A detailed discussion and definition of the tuning measure is provided in section~\ref{sec:models} and appendix~\ref{sec:FT}. 

We find that with the $m_h=125$ GeV constraint, the least-tuned spectra can be accessible at the 14 TeV LHC, but are generally  beyond the reach at 8 TeV. This suggests an intriguing possibility: that the failure to find superpartners so far at the LHC could actually be a consequence of $m_h=125$ GeV, rather than a separate issue.
 
 Here is a synopsis of our results for each kind of model:
 \begin{itemize}
 \item We confirm that the type I Higgs models  (some of which have been studied before in \cite{Jelinski:2011xe,Kang:2012ra,CraigShih,Albaid:2012qk}) are indeed fine-tuned at the level of $\sim 10^4$ because of the little $A/m_H^2$ problem. The spectra of the best points across the different models are rather similar.  However, the prospects for observing these models at the LHC are rather pessimistic because all of the colored objects tend to be quite heavy.  Some spectra do possess a light wino which could be produced at the LHC, but this is the exception rather than the norm. These spectra all exhibit slepton co-NLSPs with roughly 250-450 GeV masses -- within the range that an ILC could discover.
 
 \item On the other hand, in the type I squark models, there is no little $A/m_H^2$ problem, so these models are considerably less fine tuned.  To the best of our knowledge, these models have never been studied in detail before. All of these models possess points of relatively low tuning, with $\Delta_{FT}\sim 10^3$.  However, as these models are not manifestly MFV, they can in principle conflict with flavor physics constraints from precision experiment.  For the purposes of this work, we adopt an agnostic stance toward flavor (e.g.\ we assume perfect alignment) and simply aim to show that the type I squark models are promising from a tuning point of view.  Discussions of flavor physics in these models will be deferred for future work \cite{paperwithArun}.  The LHC phenomenology of these models is more promising.  Often, there is an accessible stop (and sbottom in $Q_3$ models) with slepton co-NLSPs, sometimes with a bino between the two states.  Alternatively, some of the regions of lower tuning have gluinos and squarks light enough to be produced at 14 TeV.
 
\item There are five type II models with mixing, and they have been studied in various guises in \cite{ChackoPonton,Shadmi:2011hs,Evans:2011bea,Jelinski:2011xe,Evans:2012hg,Albaid:2012qk,Abdullah:2012tq,Perez:2012mj,Endo:2012rd}. In three of these terms the interaction is top-Yukawa-like ($QU\Phi$, $H_u U \Phi$ and $H_u Q \Phi$), and in two of them it is bottom-Yukawa-like ($Q D \Phi$ and $Q H_d \Phi$). The latter receive no appreciable influence from mixing, but they are more constrained by tachyons and/or EWSB, so they manifest with poor tuning.   Conversely, the top-Yukawa-like terms are especially interesting as $A_t$ is enhanced by receiving contributions from two sources (\ref{eq:Atsum}).  This enhancement is so effective that the mixed $H_u U \Phi$ model contains the points with the least tuning out of any model studied in this work.  
Though inferior to the best regions of all type I squark models, the $H_u Q \Phi$ and $QU\Phi$ models have regions with lower tuning than the type I Higgs models and the rest of the type II models.  The regions of least tuning in the $QU\phi_{H_u}$ model are experimentally excluded by existing LHC searches because they have light $\order{1 \mbox{ TeV}}$ gluinos and first-generation squarks.  The other top-Yukawa-like models have spectra with gluinos and first generation squarks which will be accessible at 14 TeV running.  Other production avenues, especially promising at an ILC, include very light slepton co-NLSPs and light Higgsinos entering at several hundred GeV.

\item Most of the type II models without mixing (see \cite{Kim:2005qb,Joaquim:2006mn,Albaid:2012qk,Kim:2012vz} for some examples) are extremely tuned.  These models tend to either suffer from the same issues as the type I Higgs models, or introduce additional tachyons which place the model in tiny corners of viable parameter space.   However, the $U_3 D_3 \Phi_{D}$ model has regions of slightly lower tuning.  Additionally, these models depart from the slepton co-NLSP phenomenology of the type I models and will often have a bino NLSP.  Gluinos accessible at 14 TeV could lead to exciting phenomenology. 

 \end{itemize}
 
Our paper is outlined as follows:  In section~\ref{sec:soft}, we discuss the complications which arise when using analytic continuation in superspace to derive soft parameters in models with MSSM-messenger mixing.  We then present a new, completely general framework for deriving messenger scale soft parameters in the presence of any MSSM-messenger interactions.  In section~\ref{sec:models}, we catalog and survey the parameter space of the 31 possible couplings.  Details are provided in a series of subsections for type I Higgs, type I squark, type II mixed and type II unmixed models.  We discuss the phenomenology of the models with the least tuning in section~\ref{sec:pheno}.  Section~\ref{sec:conclusions} contains a summary and discussion of future directions.  We devote appendix~\ref{sec:validation} to validating the formulas of section~\ref{sec:soft} through multiple methods.  The fine-tuning measure utilized throughout this work is detailed in appendix~\ref{sec:FT}.

{\bf Note added:} while this paper was in preparation, \cite{Byakti:2013ti} appeared which overlaps partially with section 3 of our work. This paper also creates a catalog of the different models of MSSM-messenger interactions for large $A_t$. We note that their classification scheme differs from ours; our catalog contains several additional couplings absent from their study, namely $Q\phi_{Q}\phi_{\bar D}$, $U\phi_{Q}\phi_{H_u}$, $U\phi_{E}\phi_{\bar D}$, $QQ\phi_{\bar D}$, as well as all couplings containing an $SU(5)$ adjoint, $\phi_{24}$; and their formulas, being based on those of \cite{ChackoPonton}, neglect a proper treatment of MSSM-messenger mixing.

\section{A New Calculation of the Soft Spectrum}
\label{sec:soft}
\subsection{Problems with the existing derivation}

In this section, we present a new method for calculating the soft spectrum induced by MSSM-messenger interactions via analytic continuation in superspace. The current state of the art (prior to this work) are the general formulas contained in \cite{ChackoPonton}. Those formulas express the soft masses in terms of the beta functions and anomalous dimensions above and below the messenger scale. They are meant to be completely general, but are prone to misapplication whenever there is mixing between messenger and MSSM fields. 

To understand the issues with the derivation of \cite{ChackoPonton}, we need to first review some aspects of wavefunction renormalization in the presence of operator mixing in SUSY theories. Let us define the theory above the messenger scale to be:
\begin{eqnarray}
\label{eq:Kabove}
\begin{aligned}
K &= Z_{ij}^+(t)\Phi_i^\dagger \Phi_j \\
W &={1\over6}\lambda_{ijk}\Phi_i\Phi_j\Phi_k+{1\over2}\kappa_{ij}X\Phi_i\Phi_j
\end{aligned}
\end{eqnarray}
Here $t\equiv\log \mu$ is the RG scale; since we are in the holomorphic scheme, the superpotential couplings do not run. The indices $i,\,j,\,\dots$ run over all messenger and MSSM fields (transforming in SM irreps).  We allow for the possibility that any of the fields can mix with any of the others by taking the wavefunctions to be a general Hermitian matrix $Z_{ij}^+$. 

In the holomorphic scheme, passing through the messenger scale is trivial, and the wavefunctions for the MSSM fields remain continuous. Below the messenger scale, the theory is of the same form, but without any messenger fields:
\begin{eqnarray}
\label{Kbelow}
\begin{aligned}
K &= Z_{ab}^-(t;M)\Phi_a^\dagger \Phi_b \\
W &={1\over6}\lambda_{abc}\Phi_a\Phi_b\Phi_c
\end{aligned}
\end{eqnarray}
where now the $a,\,b,\,\dots$ indices are summed only over MSSM fields only. Below the messenger scale, $Z^-_{ab}$ depends on $M$ through its boundary conditions, so we denote this by $Z^-_{ab}(t;M)$.

The soft parameters are given by analytically continuing $M\to \sqrt{X^* X}$ and substituting $\langle X\rangle = M+\theta^2 F$ into
\begin{equation}
\int d^4\theta\, Z_{ab}^-(t;|X|) \Phi_a^\dagger \Phi_b  \supset F\, (\partial_X Z_{ab}^-) F_{\Phi_a}^\dagger \Phi_b + \hc + |F|^2( \partial_{X}\partial_{X^*} Z_{ab}^- )\Phi_a^\dagger\Phi_b
\end{equation}
so, to leading order:
\bea
\label{eq:Amsq}
A_{ab} &= F \partial_X Z_{ab}^- \\
m_{ab}^2 &= -|F|^2 \partial_{X}\partial_{X^*}Z_{ab}^- +|F|^2 \partial_{X^*} Z_{ac}^- \partial_X Z_{cb}^- 
\eea
with the derivatives evaluated at $t=\log |X|$.  The second term in $m_{ab}^2$ comes from integrating out the $F$-components of $\Phi_c$.  Note that because these expressions exist below the messenger scale, all indices correspond to MSSM fields only.

Finally, the derivatives of the wavefunctions can be obtained from the integral expressions:
\begin{equation}
\label{eq:Zasintegral}
Z_{ab}^-(t;M) = \int_{\log M}^{t}dt' \, {d Z_{ab}^-\over dt'}   + \int_{\log \Lambda_{\rm UV}}^{\log M}dt' \, {d Z_{ab}^+\over dt'}
\end{equation}
using the relation between the wavefunctions and the anomalous dimensions\footnote{We note that \cite{ChackoPonton} used an equation for wavefunction renormalization, ${dZ\over dt} = \gamma Z$, which is incorrect whenever $V$ and $\gamma$ do not commute.  From this starting point, \cite{ChackoPonton} derive a formula for $ m_{ab}^2$ that contains a spurious extra term which goes as the commutator of $\gamma$ above and below the messenger scale.  This must be incorrect, because it is not Hermitian. However, it drops out of the Lagrangian when sandwiched between $\Phi_a^\dagger\Phi_b$. }
\begin{equation}
\label{eq:dZdt}
Z = V^\dagger V,\qquad {dV\over dt} = -\gamma V \qquad \Rightarrow\qquad {dZ\over dt}=-2V^\dagger \gamma V
\end{equation}
Here $\gamma$ is the matrix of anomalous dimensions as a function of the non-holomorphic running couplings, which we will denote by $\tilde\lambda(t)$. These are related to the holomorphic couplings via
\begin{equation}
\label{eq:lambdarel}
\tilde\lambda_{ijk}(t)=\lambda_{i'j'k'}(V^{-1})_{i'i}(V^{-1})_{j'j}(V^{-1})_{k'k}
\end{equation}
One can check that with these equations, the beta function for $\tilde\lambda_{ijk}(t)$ is the standard one given in e.g.\ \cite{MartinVaughn}. 

The problem in the derivation of \cite{ChackoPonton} arises at this stage. Chacko \& Ponton view (\ref{eq:dZdt}) as an equation which determines ${dZ\over dt}$ in terms of the non-holomorphic running couplings $\tilde\lambda_{ijk}(t)$. These are defined in the non-holomorphic basis where the K\"ahler potential is canonical. If these were the only running couplings, then the derivation of \cite{ChackoPonton} could be safely applied. However, this is not the case: in addition to $\tilde\lambda_{ijk}$ running, the non-holomorphic couplings $\tilde\kappa_{ij}$ also run. That is, analogous to (\ref{eq:lambdarel}), we also have:
\begin{equation}
\label{eq:kapparel}
\tilde\kappa_{ij}(t) = \kappa_{i'j'}(V^{-1})_{i'i}(V^{-1})_{j'j}
\end{equation}
Thus, the theory in the non-holomorphic basis, where the derivation of \cite{ChackoPonton} takes place, is actually:
\bea
\label{eq:Kabovenonhol}
K &= \tilde\Phi_i^\dagger \tilde\Phi_j \\
W &={1\over6}\tilde\lambda_{ijk}(t)\tilde\Phi_i\tilde\Phi_j\tilde\Phi_k+{1\over2}\tilde\kappa_{ij}(t)X\tilde\Phi_i\tilde\Phi_j
\eea
Even with $\tilde\kappa_{ij}(t)$ entirely aligned with the messenger directions in the far UV, the presence of MSSM-messenger mixing will cause it to become nonzero in the MSSM direction by the time we reach the messenger threshold.  Now, integrating out the messengers will introduce some dependence on the MSSM fields, schematically: 
\begin{equation}
\tilde\Phi_i \,\,\,\longrightarrow\,\,\, C_{ia}\tilde\Phi_a
\end{equation}
Substituting this back into (\ref{eq:Kabovenonhol}), we obtain the theory below the messenger scale where the effective K\"ahler potential for the MSSM fields shifts discontinuously, 
\begin{equation}
\label{eq:Kshift}
K\to K+\tilde\Phi^\dagger C^\dagger C\tilde\Phi
\end{equation}
This potential is neither canonical nor continuous through the messenger threshold.

To summarize, the problem with applying the derivation of \cite{ChackoPonton} in the presence of MSSM-messenger mixing is that it relies on the non-holomorphic scheme where the K\"ahler potential is canonical and the superpotential couplings run. However, it also assumes that the wavefunctions of the MSSM fields are continuous through the messenger threshold. Together, these assumptions prove incompatible with the standard beta functions for the non-holomorphic couplings. 

The discontinuity in the wavefunctions (\ref{eq:Kshift}) amounts to an additional contribution to the soft masses in the presence of MSSM-messenger mixing, which is missed in the formulas of \cite{ChackoPonton}. One could attempt to take this into account by fleshing out the line of reasoning above, as done in appendix~\ref{sec:validation} for a particular model. Alternatively, one can perform an extra unitary rotation to undo the effect of the RG (\ref{eq:kapparel}) and to eliminate the extra contribution (\ref{eq:Kshift}). While this latter method yields correct results (it was done for a specific model in \cite{Evans:2012hg}), a general formula derived from this procedure does not exist. Furthermore, the price one pays in this approach is that the anomalous dimensions and beta functions are no longer given by standard formulas; in particular the matrix of anomalous dimensions is no longer Hermitian.

\subsection{A fresh approach to the calculation}

In this paper, we will take a fresh approach to the problem of computing wavefunction renormalization, one that will overcome the problems discussed above. As reviewed in the previous subsection, in the standard implementations of analytic continuation, one first puts the K\"ahler potential into a canonical form, solves the beta function equations for the non-holomorphic couplings, substitutes these into formulas for anomalous dimensions, and integrates these to get the wavefunctions $Z$ which were canonicalized in the first step. This methodology seems rather convoluted, since the running of the non-holomorphic couplings is nothing other than wavefunction renormalization.  Wouldn't it be conceptually simpler to {\it stay in the holomorphic basis} and directly integrate a single differential equation for $Z$, rather than integrating two differential equations which are really expressions of the same underlying physics?

All that is required is to view (\ref{eq:dZdt}) as an equation for $Z$ itself, and not as an equation which determines $Z$ in terms of the non-holomorphic couplings. At one-loop,  the anomalous dimensions are given by:
\bea
\label{eq:gammaij}
\gamma_{ij} &= {1\over16\pi^2}\left({1\over2}d_i^{k\ell}\tilde\lambda_{ik\ell}^*\tilde\lambda_{jk\ell}-2c_r^i\delta_{ij}g_r^2\right)\\
 & = {1\over16\pi^2}V^{-1\dagger}_{i i'} \left({1\over2}d_{i'}^{k\ell} \lambda^*_{i'k\ell} Z_{k m }^{-1*}Z_{\ell n}^{-1*}\lambda_{j'm n } -2c_r^{i'} Z_{i'j'}g_r^2\right)V^{-1}_{j'j}
\eea
Here $d_i^{k\ell}$ is a standard multiplicity factor present in the one-loop anomalous dimensions; roughly speaking it counts the number of fields of type $k$ and $\ell$ that can talk to field $i$ through the interactions. Concrete examples of $d_i^{k\ell}$s will be given in later sections. In the second line of (\ref{eq:gammaij}), we used the facts that $d_i^{k\ell}$ is a function of only the gauge representations of $i$, $k$ and $\ell$, and that $Z_{ij}$ and $V_{ij}$ can only connect two fields in the same representation. Substituting (\ref{eq:gammaij}) into (\ref{eq:dZdt}), this becomes
\begin{equation}
\label{eq:dZdtexplicit}
{d Z_{ij}\over dt} =G_{ij}[Z(t);\lambda,g(t)] \equiv  -{1\over8\pi^2}  \left({1\over2}d_{i}^{k\ell} \lambda^*_{ik\ell} Z_{k m }^{-1*}Z_{\ell n}^{-1*}\lambda_{j m n } -2c_r^{i } Z_{i j}g_r^2\right)
\end{equation}
The $V$'s have disappeared, and we have the desired form of the differential equation for $Z$ given in terms of $Z$ itself. 

It remains to compute the first and second derivatives of $Z_{ab}^-$ with respect to $X$ and $X^*$. Here we can follow essentially the same  steps as in \cite{ChackoPonton}. Repeatedly using  (\ref{eq:Zasintegral}) and (\ref{eq:dZdtexplicit}), we arrive at:
\bea
\label{eq:dZdlogXfinal}
{\partial Z_{ab}^-(\log\mu;|X|)\over \partial X}\Big|_{\mu=|X|}  &={1\over2X} \Delta G_{ab}\\
{\partial^2Z_{ab}^-(\log\mu;|X|)\over \partial X \partial X^*}   \Big|_{\mu=|X|} &= {1\over 4|X|^2} \left( \Delta\left( {\partial G_{ab}\over\partial Z_{ij}}\right)G^+_{ij} - {\partial G^-_{ab}\over\partial Z^-_{ij}} \Delta G_{ij} + \Delta\left( {\partial G_{ab}\over\partial g_r}\right) \beta_{g_r}^+ -  {\partial G^-_{ab}\over\partial g_r^-} \Delta\beta_{g_r}\right)
\eea
where $\Delta(\dots)$ means the discontinuity of $(\dots)$ across $t=\log |X|$. These results are clearly analogous to the formulas in \cite{ChackoPonton}, but they are more broadly applicable. In particular, there is no complication in applying this formula to models with MSSM-messenger mixing. Substituting the explicit formula for $G_{ij}$ given in (\ref{eq:dZdtexplicit}) into (\ref{eq:dZdlogXfinal}) and then into (\ref{eq:Amsq}),  we obtain at leading loop order,
\bea
\label{eq:Amsqfinal}
A_{ab} &  =  -{1\over 32 \pi^2}d_{a}^{ij} \Delta \left( \lambda^*_{a ij} \lambda_{b i j } \right)\Lambda \\
\delta m_{ab}^2 &=\frac{1}{256\pi^4} \lp\frac 12 d_a^{ik}d_i^{\ell m}\lp \Delta  \lp \lambda_{aik}^* \lambda_{bj k}\rp \lp \lambda_{i\ell m} \lambda_{j \ell m}^*\rp^+
 - \lp\lambda_{aik}^* \lambda_{bj k}\rp^- \Delta\lp \lambda_{i\ell m} \lambda_{j \ell m}^* \rp \rp
 \right. \\ & \left. + \frac 14 d_a^{ij} d_b^{k\ell} \Delta\lp \lambda_{a ij}^* \lambda_{cij}\rp\Delta\lp\lambda_{ck\ell}^* \lambda_{bk\ell}\rp
 - 
 d_a^{ij} C_r^{aij}g_r^2  \Delta\lp \lambda_{aij}^* \lambda_{bij}\rp   \rp \Lambda^2
\eea
where $\Lambda = F/M$, and $C_r^{ijk} = c_r^i+c_r^j+c_r^k$ is the sum of the quadratic Casimirs of each field interacting through $\lambda_{ijk}$.  In this expression, we have not bothered to write the usual GMSB term (hence the $\delta$ in front of $m_{ab}^2$), which comes from the last two terms in the second line of (\ref{eq:dZdlogXfinal}). All indices are summed over except for $a$ and $b$.

This is our final, general result for the one-loop $A$-terms and two-loop mass-squared terms induced by MSSM-messenger interactions. On top of the standard GMSB contribution to the soft masses, one must add to this expression an additional term appearing at one-loop, but suppressed by $\frac{\Lambda^2}{M^2}$ \cite{CraigShih}:
\beq
\label{eq:1loop}
\delta m^2_{ab,{1-loop}} = - \frac { h\lp \Lambda/M \rp }{96\pi^2} d_{a}^{ij} \Delta \lp \lambda_{aij}^*\lambda_{bij} \rp\frac{\Lambda^4}{M^2}
\eeq
where 
\beq
\label{eq:h1loop}
h(x) =\frac{3}{x^4} \Big( (x-2) \ln(1-x) - (x+2) \ln(1+x)  \Big) = 1+ \frac 45 x^2 + \order{x^4}
\eeq
Using these formulas, one can derive correct expressions for the soft masses and $A$-terms for any model with any number of type I and type II MSSM-messenger interactions, including those with mixing between any and all sectors.  However, in the following subsections, we present specific simplified formulas for models containing type I couplings only and type II couplings only.  These formulas are used throughout this work. 

\subsection{Formulas for type I models}

In the previous section, we derived the most general formulas for the MSSM soft masses, including the possibility of arbitrary mixing between MSSM and/or messenger fields. Now, we would like to specialize to models which involve only type I or type II couplings, beginning with the type I case. 

To improve the readability of the formulas, it will be convenient to introduce messenger-only indices $A,\,B,\,\dots$. The indices $a,\,b,\,\dots$ will continue to run over MSSM fields only; and $i,\,j,\,\dots$ will continue to run over all fields. Finally we will denote MSSM-messenger interactions with $\lambda$, but MSSM-only interactions (the usual Yukawa couplings) with $y$.

In the type I models, the interaction is of the MSSM-messenger-messenger type:
\begin{equation}
W ={1\over2} \lambda_{a B C} \Phi_a \Phi_{B}\Phi_{C}
\end{equation}
In these models, one can always impose messenger parity, so that MSSM-messenger mixing does not occur. Specializing (\ref{eq:Amsqfinal}) to the type I case, we obtain:
\bea
\label{eq:typeIgen}
A_{ab} &=  -{1\over32\pi^2}d_{a}^{BC} \lambda_{aBC}^*\lambda_{bBC} \Lambda\\
\delta m_{ab}^2 &= {1\over 256\pi^4}\Bigg( d_a^{ B C}d_{B}^{c D}\lambda_{a B C}^*\lambda_{b C E} \lambda_{c B D}\lambda_{c DE}^* + {1\over4}d_{a}^{B C }d_{b}^{DE} \lambda_{a B C}^* \lambda_{c B C } \lambda_{c DE}^* \lambda_{bDE}\\
&\qquad - {1\over2}d_a^{cd}d_c^{BC}y_{acd}^* y_{b d e}\lambda_{c BC}\lambda_{e BC}^* -d_{a}^{BC}C_r^{aBC}g_r^2  \lambda_{aBC}^*\lambda_{bBC}\Bigg)\Lambda^2
\eea
If we further specialize to the case of no MSSM-MSSM mixing and no messenger-messenger mixing (e.g.\ only a single coupling between MSSM and messenger sectors), then this becomes
\bea
\label{eq:typeI}
A_a &=-{1\over32\pi^2}d_{a}^{BC} |\lambda_{a BC}|^2 \Lambda\\
\delta m_{a}^2 &= {1\over 256\pi^4}\Bigg( d_a^{BC}d_{B}^{cD}|\lambda_{aBC}|^2 | \lambda_{c B D}|^2  + {1\over4}d_{a}^{BC}d_{a}^{DE} |\lambda_{aBC}|^2 |\lambda_{aDE}|^2\\
&\qquad - {1\over2}d_a^{cd}d_c^{BC}|y_{acd}|^2|\lambda_{c BC}|^2-d_{a}^{BC} C_r^{aBC} g_r^2  |\lambda_{aBC}|^2  \Bigg)\Lambda^2
\eea
which agrees exactly with the formulas given in appendix A of \cite{CraigShih}.

\subsection{Formulas for type II models}
In our framework, it is clear that the type II models with mixing are really no different than the type II models without mixing.  When a particular model does not have any MSSM-messenger mixing, some of the terms in $\delta m_{ij}^2$ are simply zero.  Nothing special needs to be done and the mixing is fully accounted for by the formulas without requiring any further treatment.   In all type II models, the interaction is of the MSSM-MSSM-messenger type:
\begin{equation}
W ={1\over2} \lambda_{abC} \Phi_a \Phi_b\Phi_C
\end{equation}
Specializing to this case, the soft SUSY breaking terms are now given by:
\bea
\label{eq:typeIIgen}
A_{ab} &= -{1\over16\pi^2}d_{a}^{cB}\lambda_{a c B}^*\lambda_{b c B}\Lambda\\
\delta m_{ab}^2 &= {1\over256\pi^4}\Bigg( 
{1\over2}d_a^{c B}d_{B}^{de} \lambda_{a c B}^*\lambda _{b c C}\lambda_{de B}\lambda_{d e C}^* 
+ d_a^{cB}d_c^{d C}\lambda_{a c B}^*\lambda_{b e B}\lambda_{c d C}\lambda_{ d e C}^* \\
& + d_{a}^{c B}d_{b}^{d C}\lambda_{a c B}^*\lambda_{c e B}\lambda_{d e C}^*\lambda_{b d C}
-d_a^{c d}d_c^{fB} y_{a c d}^* y_{b d e }\lambda_{c f B}\lambda_{e f B}^* + {1\over2}d_a^{c B}d_c^{ef}y_{c ef}y_{d ef}^* \lambda_{a c B}^*\lambda_{bd B} \\
&  + \frac 12 d_a^{cd}d_c^{ef} y_{acd}^* y_{cef}\lambda_{b d B}\lambda_{ef B}^*  
+ \frac 12 d_a^{cB}d_B^{ef} \lambda_{a c B}^* \lambda_{ef B} y_{bcd} y_{def}^* 
 -2d_{a}^{c B}C_r^{acB} g_r^2 \lambda_{a c B}^*\lambda_{b c B }
\Bigg)\Lambda^2
\eea
The first two terms in the last line are the additional contributions which arise from couplings with MSSM-messenger mixing. It is easy to see that these indeed vanish if there is no MSSM-messenger mixing, since in that case $\lambda_{efB}$ and $y_{cef}$ cannot be simultaneously nonzero.  It can be checked that naively substituting the standard beta functions and anomalous dimensions into the formulas of \cite{ChackoPonton} misses precisely these extra terms. 

If we assume there is no MSSM-MSSM or messenger-messenger mixing, then (\ref{eq:typeIIgen}) becomes:
\bea
\label{eq:typeIIsimp}
A_{a} &= -{1\over16\pi^2}d_{a}^{c B} \abs{\lambda_{acB}}^2\Lambda\\
\delta m_{a}^2 &= {1\over256\pi^4}\Bigg( 
{1\over2}d_a^{ c B }d_{B}^{de} |\lambda_{a c B}|^2 |\lambda_{de B}|^2
+ d_a^{c B}d_c^{dC}|\lambda_{acB}|^2 |\lambda_{c dC}|^2 \\
& + d_{a}^{cB}d_{a}^{dC}|\lambda_{acB}|^2 |\lambda_{adC}|^2
  -d_a^{cd}d_c^{f B} | y_{acd}|^2 |\lambda_{c f B}|^2 + {1\over2}d_a^{c B}d_c^{ef} |y_{cef}|^2 |\lambda_{acB}|^2\\
& + \frac 12  d_a^{cd}d_c^{ef} y_{acd}^* y_{cef}\lambda_{a dB}\lambda_{efB}^*  + \frac 12 d_a^{cB}d_{B}^{ef} \lambda_{ac B}^*\lambda_{efB}y_{acd} y_{d ef}^* 
 -2d_{a}^{cB}C_r^{acB}g_r^2  |\lambda_{a cB}|^2
\Bigg)\Lambda^2
\eea
where, as before, the first two terms in the last line will vanish in the absence of MSSM-messenger mixing.

\section{Models}
\label{sec:models}

From the general considerations of the previous section, we now turn our focus to surveying the different types of MSSM-messenger interactions. To produce our catalog of models, we impose a few conditions. First, we require that the messengers come in complete, vector-like $SU(5)$ representations and that the SM gauge couplings remain perturbative up to the GUT scale. The relevant $SU(5)$ representations and their $SU(3)\times SU(2)\times U(1)$ decompositions are:
\bea
\label{eq:su5reps}
\phi_1 &\to  \lp {\bf 1}, {\bf 1}\rp_0\\
\phi_5\oplus\phi_{\bar 5} &\to \lp {\bf 3},{\bf 1}\rp_{-\frac13} \oplus \lp {\bf 1},{\bf 2}\rp_{\frac 12} \oplus \lp {\bf \bar 3},{\bf 1}\rp_{\frac13} \oplus \lp {\bf 1},{\bf 2}\rp_{-\frac 12}  \\
\phi_{10}\oplus \phi_{\bar{10}} &\to \lp {\bf 3},{\bf 2}\rp_{\frac16} \oplus \lp {\bf \bar{3}},{\bf 1}\rp_{-\frac 23} \oplus \lp {\bf 1},{\bf 1}\rp_{1}\oplus \lp {\bf \bar 3},{\bf 2}\rp_{-\frac 16} \oplus \lp {\bf 3},{\bf 1}\rp_{\frac 23} \oplus \lp {\bf 1},{\bf 1}\rp_{-1}  \\
\phi_{24} &\to \lp {\bf 8},{\bf 1}\rp_{0} \oplus \lp {\bf 1},{\bf 3}\rp_{0}  \oplus \lp {\bf 1},{\bf 1}\rp_{0} \oplus \lp {\bf 3},{\bf 2}\rp_{-\frac56} \oplus \lp {\bf \bar{3}},{\bf 2}\rp_{\frac56}
\eea
In order to keep the gauge couplings perturbative up to the GUT scale, the total messenger contribution to the beta function must satisfy $|\Delta b|\leq 6$. $\phi_1$, $\phi_5\oplus\phi_{\bar 5}$, $\phi_{10}\oplus\phi_{\bar{10}}$ and $\phi_{24}$ contribute 0, -1, -3 and -5 to the $SU(5)$ beta function, respectively.  

Second, we will only consider models where a single superpotential coupling $\lambda$ is turned on between the MSSM and messenger fields.  For type I models, our interaction superpotential will consist of:
\begin{equation}
\label{eq:WtypeI}
W = \lambda\, q\sum_{i=1}^{N_{m}} \phi_i\phi'_i
\end{equation}
Here, the MSSM field, $q$, must be one of either $H_u$, $Q_3$, or  $U_3$ in order to produce large $A_t$. Meanwhile, $\phi_i$, $\phi'_i$ are messenger fields transforming in $SU(3)\times SU(2)\times U(1)$ irreducible representations. In (\ref{eq:WtypeI}), we have included the possibility that the messenger number, $N_m$, may differ from one -- a viable option for type I models. Note that we have taken all couplings to the multiple pairs of messengers to be $\lambda$ for simplicity.

For type II models, the story is nearly identical. Here, our interaction superpotential will be:
\begin{equation}
\label{eq:WtypeII}
W = \lambda q q' \phi
\end{equation}
with $q$, $q'$ being MSSM fields (at least one of which must be either $H_u$, $Q_3$, or  $U_3$), and $\phi$ a messenger transforming as an SM irrep. Note that for type II models there is no option to talk to multiple messengers; the messenger number is always one for the contribution from the MSSM-messenger interactions. The messenger number for the GMSB contribution could be greater than one, but this always serves to make the model more fine-tuned (it will reduce $A_t/M_S$), so we will restrict ourselves to $N_m=1$ for type II models.

Under these constraints, there are 31 models in all -- 15 of type I and 16 of type II. These are cataloged in Table~\ref{table:models}. Each model in this table can be parametrized in a uniform way. As in \cite{CraigShih}, we will choose the parameter space to be (for simplicity, $\tan \beta=10$ throughout this work):
\begin{equation}
\left(\lambda,\,\,{\Lambda\over M},\,\,N_m,\,\,\Lambda\right)
\end{equation}
For a given value of the first three parameters, increasing $\Lambda=F/M$ simply increases the overall scale of the sparticle masses and $A$-terms (while keeping the stop mixing fixed).  Since $m_h$ increases monotonically with $\Lambda$ (by having heavier stops running in the loops),  $\Lambda$ is uniquely determined by imposing $m_h=125$ GeV.  This procedure can fail if some sparticles are tachyonic or if the basic conditions for electroweak symmetry breaking cannot be satisfied. 

In almost all of the models which we discuss, either $U_3$ or $Q_3$ does not talk to the messenger directly.  That field will run tachyonic at high values of $\lambda$ due to the leading terms in the soft masses being of the form,
\beq
\label{eq:sizestop}
\lp16 \pi^2\rp^2 \frac{m_{stop}^2}{\Lambda^2} =  -  B_1 y_t^2 \lambda^2 + B_2 g_3^4.
\eeq
where $B_1$, $B_2>0$ are the contributions from the MSSM-messenger coupling and the standard GMSB contribution, respectively.  So at a large enough value of $\lambda$, the squark becomes tachyonic. Solving for $\lambda$, this happens near, 
\beq
\label{eq:size}
\lambda_{max}\sim \sqrt{\frac {B_2 g_3^4}{B_1 y_t^2}}\approx  \sqrt{\frac {B_2}{B_1}}
\eeq
This squark tachyon ceiling appears in nearly all models, although RG running from the messenger scale perturbs the actual value of $\lambda_{max}$ (which manifests in gradual curves, as opposed to horizontal lines in $\lp\lambda,\LoM\rp$ space).

To explore each of these models, a dense grid of points in $\lambda$ and $\LoM$ is generated, with $\Lambda$ chosen to satisfy the $m_h=125$ GeV condition. We use SOFTSUSY~\cite{Allanach:2001kg} to run the spectra from the messenger scale to the TeV scale. As in \cite{CraigShih}, contour plots in the $(\lambda,\LoM)$ plane completely characterize the viable parameter space of the model. If a particular point gives a valid solution, then the fine-tuning at that point is calculated using our tuning measure.  As discussed in more detail in appendix~\ref{sec:FT}, the tuning measure is defined as
\beq
\label{eq:FTbody}
\Delta_{FT} \equiv \max \Delta_i \;\; \mbox{ where } \;\; \Delta_i \equiv \frac{\d \log m_z^2}{\d \log \Lambda_i^2},
\eeq
 where $\Lambda_i\in \{\Lambda_\lambda,\Lambda_3,\Lambda_t, \Lambda_{1-loop},\mu \}$ with $\Lambda_\lambda \equiv \lambda^2\Lambda$, $\Lambda_3 \equiv g_3^2\Lambda$, $\Lambda_t \equiv y_t^2\Lambda$ and $\Lambda_{1-loop}$ measures the variation of the one-loop term (\ref{eq:1loop}).   Each quantity is treated independently and varied separately.

  The least tuned point located in each model is cataloged in table~\ref{table:models}.  The spectra of the models with low tuning will be discussed in detail in section~\ref{sec:pheno}.
 The $U H_u \phi_Q$ type II model is the least tuned ($\Delta_{FT}\sim850$) out of all models, however, the rest of the type II models fare poorly with regards to tuning.  In general, all of the type I squark models enter with a relatively low tuning measure of $\Delta_{FT}\sim10^3$.      Many of the models involving Higgs fields have very large $M_S$ (and small $\abs{A_t}/M_S$) because they are relying on heavy stops to generate $m_h=125$, as opposed to using maximal mixing.  As these models are unable to achieve maximal mixing without substantial tuning entering elsewhere (due to the little $A/m_{H_u}^2$ problem), we make no effort to optimize the tuning in these models by scanning regions of parameter space where the MSSM-messenger contributions are small.   Details concerning the various models will be discussed in the next subsections.

\begin{table}[t]
\begin{center}
\begin{tabular}{|c|c|c|c|c|c|c|c|c|}
\hline
\# & Coupling  & $|\Delta b|$  & Best Point $\{\frac \Lambda M , \lambda \}$ & $\abs{A_t}/M_S$ & $M_{\tilde g}$ & $M_S$ & $\abs{\mu}$ & Tuning \\
\hline \hline
I.1 & $H_u \phi_{\bar5,L} \phi_{1,S}$ & $N_{m}$   & \braced{0.375,1.075} &1.98& 3222&1842 &777 & 3400 \\
I.2 & $H_u \phi_{10,Q} \phi_{10,U}$ & $3N_{m}$ & \braced{0.25,1.075} & 1.99 & 3178 & 1828 & 789&  2450 \\
I.3 & $H_u \phi_{5,\bar D} \phi_{\bar{10},\bar Q}$  & 4 & \braced{0.25,1.3} & 2.05 & 2899 & 1709 & 668 & 3200 \\
I.4 & $H_u \phi_{5,\bar L} \phi_{\bar{10},\bar E}$   & 4 &  \braced{0.125,0.95}&0.58&11134&8993&2264&4050 \\
I.5 & $H_u \phi_{\bar5,L} \phi_{24,S}$ & 6  &  \braced{0.225,1.000}&0.54&13290&9785&3408&3850 \\
I.6 & $H_u  \phi_{\bar5,L} \phi_{24,W}$ & 6  &\braced{0.15,1.025}&0.67&11835&8637&3259&3410 \\
I.7 & $H_u \phi_{\bar5,D} \phi_{24,X}$  & 6  & \braced{0.3,1.425} & 2.04& 3020 & 1743 & 576 & 3500  \\
\hline
I.8 & $Q \phi_{\bar{10},\bar Q} \phi_{1,S}$  & $3N_{m}$   & \braced{0.534,1.5} & 2.82 & 4336 &1274 & 2056 & 1015   \\
I.9 & $Q \phi_{\bar5,D}  \phi_{\bar5,L}$  & $N_{m}$   & \braced{0.353,0.858} &2.67&4247&1342&2058&1015  \\
I.10  & $Q \phi_{10,U} \phi_{5,H_u}$  &  $4$   & \braced{0.51,1.788}&2.65&4040&1318&2301&1275  \\
I.11 & $Q \phi_{10,Q} \phi_{5,\bar D}$ &  $4$ & \braced{0.378,1.245}&2.76&4020&1257&2292&1260  \\
\hline
I.12 & $U \phi_{\bar{10},\bar U} \phi_{1,S}$  & $3N_{m}$   & \braced{0.476,1.622}&2.62&3815&1347&2070&1030  \\
I.13 & $U \phi_{\bar5,D}  \phi_{\bar5,D}$  & $2N_{m}$   & \braced{0.301,0.908}&2.91&3829&1199&2061 &1020  \\
I.14 & $U \phi_{10,Q} \phi_{5,H_u}$  & 4  & \braced{0.37,1.352}&2.81&3575&1220&2312&1285 \\
I.15 & $U \phi_{10,E} \phi_{5,\bar D}$  & 4  & \braced{0.51,1.972}&2.63&3526&1312&2310&1280 \\
\hline \hline
II.1 & $Q U \phi_{5,H_u}$  & 1   & \braced{0.55,1.64}&2.02&769&1965&2738&1800 \\ 
II.2 & $U H_u \phi_{10,Q}$   & 3   & \braced{0.009,1.067}&2.14&2203&1628&543&850  \\ 
II.3 & $Q H_u \phi_{10,U}$  & 3   & \braced{0.269,1.05} &2.27&2514&1458&439 & 1500 \\ 
II.4 & $Q D \phi_{\bar5,H_d}$  & 1  & \braced{0.37,1.2} &1.78&2597&1829&3553& 3020 \\
II.5 & $QH_d \phi_{\bar5,D}$   & 1  & \braced{0.15,1.19}&1.45&2497&2108&3773&6050 \\
\hline
II.6 & $Q Q \phi_{5,\bar D}$  &  1  & \braced{0.45,0.1}&0.22&7943&9870&3610&5000 \\ 
II.7 & $U D \phi_{\bar5,D}$    &  1  & \braced{0.21,1.26}&2.34&1374&1334&2998&2150\\ 
II.8 & $Q L \phi_{\bar5,D}$   & 1  & \braced{0.14,1.2} &1.51&1501&1204&2203 & 3700 \\ 
II.9 & $U E \phi_{5,\bar D}$ & 1 & \braced{0.445,1.46}&1.89&2004&1750&3373&2730 \\ 
II.10 & $H_u D \phi_{24,X}$  & 5  & \braced{0.42,1.45} & 2.13&2943&1649&282& 3500 \\
\hline
II.11 & $H_u L \phi_{1,S}$  & $1^*$ & \braced{0.15,0.675}&0.54&7103&8166&3714&4930 \\ 
II.12 & $H_u L \phi_{24,S}$  &  5  & \braced{0.296,0.96}&0.53&12629&9660&3333&3780  \\
II.13 & $H_u L \phi_{24,W}$  & 5  & \braced{0.212,0.96}&0.65&11487&8710&3687&3380 \\
II.14 & $H_uH_d \phi_{1,S}$   & $1^*$  & \braced{0.125,0.675}&0.55&7049&8051&3255&5000 \\
II.15 & $H_u H_d \phi_{24,S}$   & 5  & \braced{0.20,1.00}&0.57&12047&9213&1628&4220\\
II.16 & $H_u H_d \phi_{24,W}$   & 5 & \braced{0.2,0.946}&0.64&11571&8789&3665&3460  \\
\hline
\end{tabular}
\caption{All possible marginal MSSM-messenger couplings compatible with a perturbative $SU(5)$ framework are tabulated here.  The point with the least tuning in each model is also presented.  The tuning measure used is defined in (\ref{eq:FTbody}) and is discussed more in Appendix~\ref{sec:FT}. Additionally, the values of $\abs{A_t}/M_S$, $M_{\go}$, $M_S$ and $\abs{\mu}$ at this least tuned point are shown.  Models with $\abs{A_t}/M_S<1$ rely on heavy stops as opposed to mixed stops.  Models II.11-13 generate large neutrino masses.  Models II.14-16 possess a $\mu/B\mu$ problem.  In the third column, $|\Delta b|$ refers to the messenger contribution to the $SU(5)$ beta function. As the singlet does not contribute to GMSB, models II.11 and II.14 are assigned an additional $\phi_5\oplus\phi_{\bar 5}$.  }
\label{table:models}
\end{center}
\end{table}

\subsection{Type I Higgs couplings}

We now survey the models in some detail, beginning with the type I Higgs models.  As discussed in \cite{CraigShih}, these models are all MFV, but they have high tuning because of the little $A/m_{H_u}^2$ problem. 

\begin{figure}[t]
\begin{center}
\includegraphics[scale=.71]{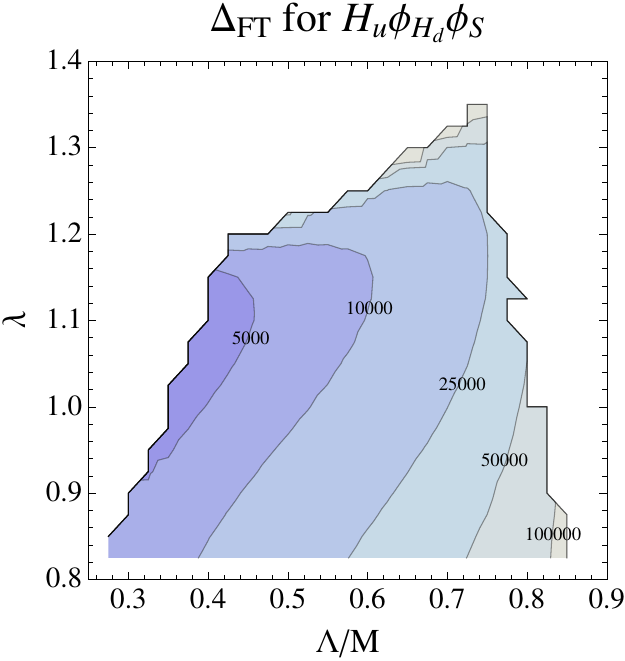}\qq
\includegraphics[scale=.69]{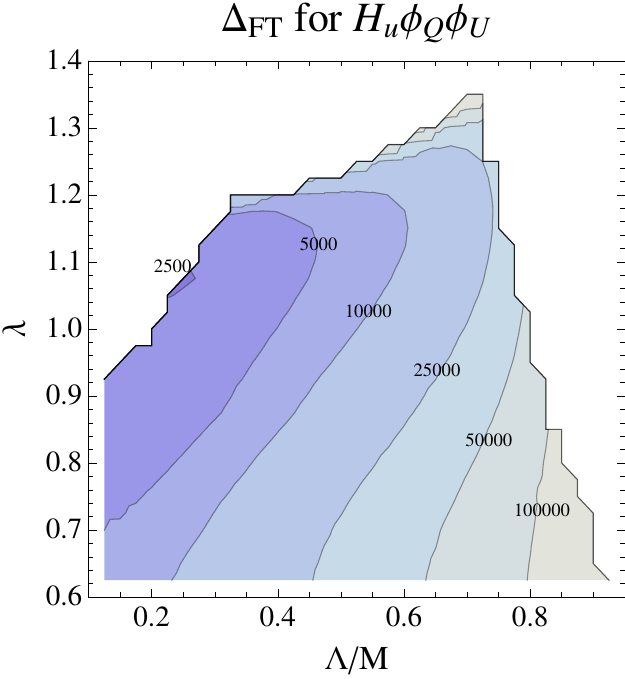}
\caption{Examples of the type I Higgs model parameter space.  Left: $H_u \phi_{\bar5,H_d} \phi_{1,S}$ with $N_{m}=6$.  Right: $H_u \phi_{10,Q} \phi_{10,U}$ with $N_{m}=2$.  In these, and all other type I Higgs models, several features emerge.  At high $\LoM$, slepton tachyons appear.  At low $\LoM$, increasing $\lambda$ causes issues with EWSB.  Moderate $\LoM$ allows for the largest $\lambda$, but this growth is truncated by a tachyonic stop, the scale of which is defined in (\ref{eq:size}).  At moderate $\lambda$ and low $\LoM$, i.e. the upper left corner of the viable region, both plots display where most type I Higgs models have their region of least tuning.}
\label{fig:HiggsType}
\end{center}
\end{figure}

With only a single MSSM-messenger coupling of the form $\lambda H_u \phi_1\phi_2$ appearing in the superpotential, the general form of the soft parameters from (\ref{eq:typeI}) can be easily specialized to the case of type I Higgs models:
\bea
A_{H_u} &= -\frac 1{16\pi^2} d_H \lambda^2  \Lambda \\
\delta m_{H_u}^2 &= \frac{1}{256\pi^4} \lp \lp d_H^2 + d_H d_{\phi}\rp \lambda^4 - 2 d_H C_r g_r^2\lambda^2  \rp \Lambda^2 - \frac{ d_H \lambda^2 h\lp\LoM\rp}{48\pi^2 } \frac{\Lambda^4}{M^2} \\
\delta m_{Q}^2 &= - \frac{1}{256\pi^4} d_H \lambda^2 y_t^2 \Lambda^2\\
\delta m_{U}^2 &=  - \frac{2}{256\pi^4} d_H \lambda^2 y_t^2 \Lambda^2
\eea
where the multiplicity factors are defined $d_H\equiv d_{H_u}^{\phi_1\phi_2}$, $d_\phi\equiv d_{\phi_1}^{H_u\phi_2}+d_{\phi_2}^{H_u\phi_1}$ and $C_r = c_r^{H_u}+c_r^{\phi_1}+c_r^{\phi_2}$ is the sum of the quadratic Casimirs.  The values in each model for $d_H$, $d_\phi$, and $C_r$ are displayed in table~\ref{table:TIHmodels}. 

\begin{table}[t]
\begin{center}
\begin{tabular}{|c|c|ccc|}
\hline
\#&Model & $d_H$ & $d_\phi$  & $C_r$ \\ 
\hline \spacer
I.1&$H_u \phi_{\bar5,H_d} \phi_{1,S}$ & $N_m$ & $3$ & $\lp\frac 3{10},\frac 32,0\rp$  \\ \spacer
I.2&$H_u \phi_{10,Q} \phi_{10,U}$  & $3 N_m$  & $3$ & $ \lp\frac {13}{30},\frac 32, \frac 83\rp$ \\ \spacer
I.3&$H_u \phi_{5,\bar D} \phi_{\bar{10},\bar Q}$ & $3$ & $3$ & $\lp \frac 7{30},\frac 32,\frac 83\rp$ \\ \spacer
I.4&$H_u \phi_{5,\bar L} \phi_{\bar{10},\bar E}$ &  $1$ & $3$ & $\lp \frac 9{10},\frac 32,0\rp$ \\ \spacer
I.5&$H_u  \phi_{\bar5,L} \phi_{24,S}$ & $1$ & $3$ & $\lp \frac 3{10},\frac 32,0 \rp$ \\ \spacer
I.6&$H_u \phi_{\bar5,L} \phi_{24,W}$ & $\frac 32$ & $\frac 52$ & $\lp \frac 3{10} ,\frac 72,0\rp$ \\ \spacer
I.7&$H_u \phi_{\bar5,D} \phi_{24,X}$ & $3$ & $3$ & $\lp \frac {19}{30} ,\frac 32,\frac 83\rp$ \\ 
\hline 
\end{tabular}
\caption{The definitions of the parameters appearing in the formulas for the type I $H_u$ models. The multiplicity factors are defined $d_H\equiv d_{H_u}^{\phi_1\phi_2}$, $d_\phi\equiv d_{\phi_1}^{H_u\phi_2}+d_{\phi_2}^{H_u\phi_1}$ and $C_r = c_r^{H_u}+c_r^{\phi_1}+c_r^{\phi_2}$ is the sum of the quadratic Casimirs.}
\label{table:TIHmodels}
\end{center}
\end{table}

Qualitatively, all of the type I Higgs models have nearly congruent parameter spaces, with three common, characteristic features emerging, see fig.~\ref{fig:HiggsType}.  These features were discussed at length in \cite{CraigShih}; let us briefly review them here. At lower $\LoM$, increasing $\lambda$ gives a very large positive contribution to $m_{H_u}^2$ which quickly causes issues with EWSB.  On the other hand, raising $\LoM$ increases the negative contribution to $m_{H_u}^2$ from the one-loop term (\ref{eq:1loop}), which makes the viable region of $\lambda$ larger.  These two features combine to form the left, slanted edge of the viable regions shown in fig.~\ref{fig:HiggsType}. Increasing $\LoM$ even further eventually caps $\lambda$ by confronting it with a stop squark tachyon and  drives $H_u$ even more negative (by raising the 1-loop term). As the model approaches $\LoM\sim1$, eventually the increasingly negative $m_{H_u}^2$ drives the sleptons tachyonic before the EWSB scale.  

In different regions of the parameter space, the tuning parameter is dominated by different quantities.  For low values of $\lambda$, the tuning is set by $\mu$ and $\Lambda_3$.  This is the case in all models, because as $\lambda\to0$, the contributions from both $\Lambda_\lambda$ and $\Lambda_{1-loop}$ vanish and $\Lambda_t$ is generally subdominant.  For moderate $\lambda$ and small $\LoM$, $\Lambda_\lambda$ is the biggest contributor to tuning. It is here that we find the lowest values of $\mu$, and hence the lowest values of the tuning parameter. Finally, for larger $\LoM$, the $\Lambda_{1-loop}$ contribution governs the tuning.  
 
Finally, let us comment on some of the differences between type I Higgs models that are apparent from table \ref{table:models}. 

\begin{itemize}

\item The first two models -- $H_u \phi_{\bar5,H_d} \phi_{1,S}$ and $H_u \phi_{10,Q} \phi_{10,U}$ -- were studied in detail in \cite{CraigShih}. For these models, different $N_m$ are possible. In type I Higgs models, increasing the messenger number within a specific model decreases the tuning for that model.   This happens because $A_t$ and $m_\st^2$ both scale as the number of messengers, thus $A_t\propto \sqrt{N_{m}} M_S$, which makes it easier to achieve $A_t\sim \sqrt 6 M_S$ for a maximal contribution to the Higgs mass. For all the other type I Higgs models, only $N_m=1$ is possible because anything greater would violate the $|\Delta b|\le 6$ constraint. 

\item Models I.4, I.5 and I.6 have small $A_t/M_S$.  These models  forfeit maximal mixing in exchange for heavy stops.  Models I.4 and I.5 are both identical to model I.1 in terms of their $A$-terms and their $\lambda$ contributions to soft masses, but in terms of their GMSB contribution to soft masses they have effective messenger number 4 and 6 respectively.  As these models are unable to achieve maximal mixing without substantial tuning entering elsewhere, we do not optimize the tuning further by scanning into regions of parameter space where the MSSM-messenger contributions are vanishing.

\item Models I.3 and I.7 both have slightly better tuning than I.4-6. This is because these models receive an enhancement from the multiplicity factor $d_{H_u}=3$.  This enhancement provides larger $A$-terms and allow significant stop mixing to be achieved.
  
  \end{itemize}

\subsection{Type I squark couplings}
\label{sec:Squarks}

In the type I Higgs models, $m_{H_u}^2$ received a large correction $+A_{H_u}^2$, leading to the little $A/m_{H_u}^2$ problem and greater fine-tuning.  Type I squark models receive an analogous correction to $m_{Q_3}^2$ or $m_{U_3}^2$, but this poses much less of a problem, as the electroweak symmetry breaking scale is only sensitive to the stop masses at loop level.  Without the little $A/m_{H_u}^2$ problem, the type I  squark models exhibit a significantly  reduced tuning with respect to the type I Higgs models. 

While the type I squark models fare well with regards to tuning, the lack of an MFV structure makes them more dangerous with regards to flavor constraints.  These constraints can be evaded, however, if the EGMSB interactions are aligned with the third generation. Obviously, a sufficiently small perturbation around perfect alignment will continue to satisfy  flavor constraints.  Precisely how small this perturbation must be and whether this alignment can be achieved naturally are interesting questions for future studies~\cite{paperwithArun}.  Regardless, while flavor appears alarming in these models, these concerns are insufficient to invalidate the models outright.

With only a single MSSM-messenger coupling of the form $\lambda Q \phi_1\phi_2$ or $\lambda U \phi_1\phi_2$ appearing in the superpotential, the general form of the soft parameters can be derived from (\ref{eq:typeI}).  For the type I $Q$-type models, we have,
\bea
A_Q &= -\frac 1{16\pi^2} d_Q \lambda^2  \Lambda \\
\delta m_{Q}^2 &=\frac{1}{256\pi^4} \lp \lp d_Q^2 + d_Q d_{\phi}\rp \lambda^4 - 2 d_Q C_r g_r^2\lambda^2  \rp \Lambda^2 - \frac{ d_Q \lambda^2 h\lp\LoM\rp}{48\pi^2 }\frac{\Lambda^4}{M^2} \\
\delta m_{H_u}^2 &= - \frac{3}{256\pi^4} d_Q \lambda^2 y_t^2 \Lambda^2 \hspace{2cm}
\delta m_{H_d}^2 =  - \frac{3}{256\pi^4} d_Q \lambda^2 y_b^2 \Lambda^2 \\
\delta m_{U}^2 &=  - \frac{2}{256\pi^4} d_Q \lambda^2 y_t^2 \Lambda^2 \hspace{2cm}
\delta m_{D}^2 =  - \frac{2}{256\pi^4} d_Q \lambda^2 y_b^2 \Lambda^2
\eea
and for the type I $U$-type models, we have,
\bea
A_U &= -\frac 1{16\pi^2} d_U \lambda^2  \Lambda \\
\delta m_{U}^2 &=\frac{1}{256\pi^4} \lp \lp d_U^2 + d_U d_{\phi}\rp \lambda^4 - 2 d_U C_r g_r^2\lambda^2  \rp \Lambda^2 - \frac{ d_U \lambda^2 h\lp\LoM\rp}{48\pi^2 }\frac{\Lambda^4}{M^2} \\
\delta m_{H_u}^2 &= - \frac{3}{256\pi^4} d_U \lambda^2 y_t^2 \Lambda^2 \hspace{2cm}
\delta m_{Q}^2 =  - \frac{1}{256\pi^4} d_U \lambda^2 y_t^2 \Lambda^2
\eea
where as in the Higgs type models, we define the multiplicity factors $d_Q\equiv d_{Q}^{\phi_1\phi_2}$ and $d_U\equiv d_{U}^{\phi_1\phi_2}$, $d_\phi\equiv d_{\phi_1}^{Q\phi_2}+d_{\phi_2}^{Q\phi_1}$ or $d_\phi\equiv d_{\phi_1}^{U\phi_2}+d_{\phi_2}^{U\phi_1}$ and $C_r = c_r^{Q,U}+c_r^{\phi_1}+c_r^{\phi_2}$ as the sum of the quadratic Casimirs.  The values of $d_{Q,U}$, $d_\phi$, and $C_r$ in each model are shown in table~\ref{table:TISmodels}.

\begin{table}[t]
\begin{center}
\begin{tabular}{|c|c|ccc|}
\hline
\#&Model & $d_Q$ & $d_\phi$  & $C_r$ \\ 
\hline \spacer
I.8&$Q \phi_{\bar{10},\bar Q} \phi_{1,S}$ &$N_m$ & $7$ & $\lp \frac 1{30},\frac 32,\frac 83 \rp$  \\ \spacer
I.9&$Q \phi_{\bar5,D}  \phi_{\bar5,L}$ & $N_m$ & $5$ & $\lp \frac7{30},\frac 32,\frac 83 \rp$ \\ \spacer
I.10&$Q \phi_{10,U} \phi_{5,H_u}$ & $1$ & $5$ & $ \lp \frac{13}{30},\frac 32,\frac 83\rp$ \\ \spacer
I.11&$Q \phi_{10,Q} \phi_{5,\bar D}$  & $2$ & $6$ & $ \lp\frac1{10},\frac 32,4 \rp$ \\
\hline 
\end{tabular}
\begin{tabular}{|c|c|ccc|}
\hline
\#&Model & $d_U$ & $d_\phi$  & $C_r$ \\ 
\hline \spacer
I.12&$U \phi_{\bar{10},\bar U} \phi_{1,S}$ & $N_m$ & $4$ & $\lp \frac8{15},0,\frac 83 \rp$ \\ \spacer
I.13&$U \phi_{\bar5,D}  \phi_{\bar5,D}$ &   $2 N_m$ & $4$ & $\lp \frac25,0,4 \rp$ \\ \spacer
I.14&$U \phi_{10,Q} \phi_{5,H_u}$ &   $2$ & $4$ & $ \lp \frac{13}{30},\frac 32,\frac 83 \rp$ \\ \spacer
I.15&$U \phi_{10,E} \phi_{5,\bar D}$ & $1$ & $4$ & $ \lp\frac{14}{15},0,\frac 83 \rp$  \\
\hline 
\end{tabular}
\caption{The definitions of the parameters appearing in the formulas for the type I squark models.  $Q$ models are on the left, $U$ models on the right.  $d_{Q(U)}\equiv d_{Q(U)}^{\phi_1\phi_2}$ is the multiplicity factor, $d_\phi\equiv d_{\phi_1}^{Q(U)\phi_2}+d_{\phi_2}^{Q(U)\phi_1}$ and $C_r = c_r^{Q(U)}+c_r^{\phi_1}+c_r^{\phi_2}$ for the type I $Q(U)$ models.}
\label{table:TISmodels}
\end{center}
\end{table}

As was the case in the type I Higgs models, increasing $N_{m}$ improves the tuning, however, here $A_t \propto N_{m}^{\frac 14} M_S$, due to one stop scaling as $m_\st^2 \propto N^2_{m}$ and the other as  $m_\st^2 \propto N_{m}$.  The $N_{m}$ enhanced models, I.8-9 and I.12-13, possess slightly lower tuning than the models which cannot capitalize on this feature.

\begin{figure}[t]
\begin{center}
\includegraphics[scale=.7]{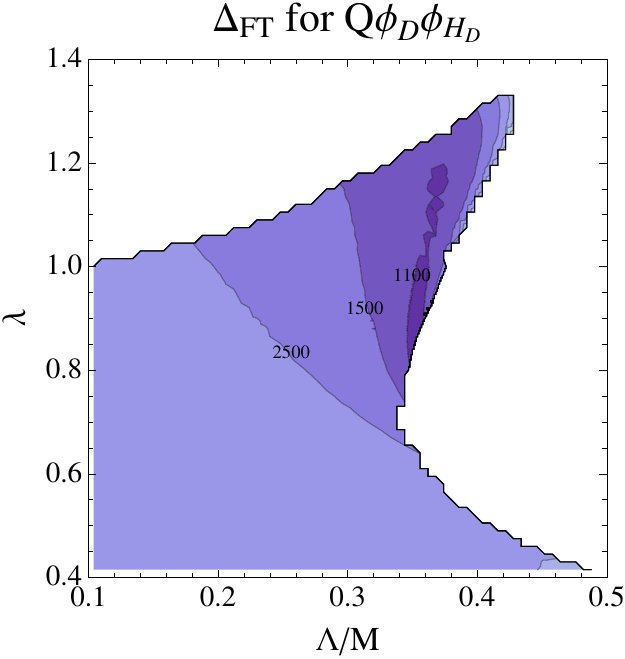}\qq
\includegraphics[scale=.7]{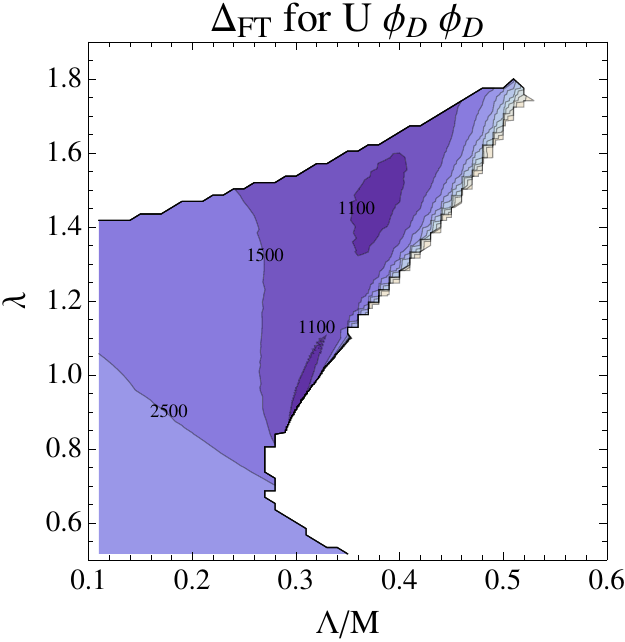}
\caption{Examples of the type I squark model parameter space.  Left:  $Q \phi_{\bar5,D}  \phi_{\bar5,L}$ with $N_{m}=6$.  Right: $U \phi_{\bar 5,D} \phi_{\bar 5,D}$ with $N_{m}=3$.  In the left (right) model, $Q_3(U_3)$ couples to the messengers, so above the horn feature, $U_3(Q_3)$ is tachyonic (\ref{eq:size}).  To the right of the horn, $Q_3(U_3)$ is tachyonic (\ref{eq:throat}).   In every type I squark model, the underside of the horn is the region of least tuning.  A second region of comparably low tuning sits in the middle of the horn for these two particular models.}
\label{fig:SquarkType}
\end{center}
\end{figure}

The parameter space of the type I squark models possesses a common, characteristic shape with two distinctive features -- the ``horn" and the ``throat" -- as shown in fig.~\ref{fig:SquarkType}.  At high $\lambda$, the squark which does not interact with the messengers (either $U_3$ or $Q_3$) is tachyonic, as discussed around (\ref{eq:sizestop}).  As $\LoM$ decreases, the messenger scale increases, so the stops have more time to run negative, thus forbidding smaller values of $\lambda$ and resulting in a curved viable region. Meanwhile, at intermediate $\lambda$ and larger $\LoM$, the squark which does interact with the messengers can go tachyonic, as its soft mass is given schematically by,
\beq
\label{eq:throatstop}
(16\pi^2)^2\frac{m_{stop}^2}{\Lambda^2} \sim B_1 \lambda^4  - B_2g_3^2 \lambda^2  + B_3 g_3^4 - \frac{8\pi^2}{ 3} B_4\lambda^2 \left({\Lambda\over M}\right)^2 h\lp\LoM\rp
\eeq  
where the $B_{i}$ are positive numbers, and the function $h(x)$ is defined in (\ref{eq:h1loop}).  The combination of stop tachyons above and below in $\lambda$ create a horn-like feature that extends into higher $\lambda$ and $\LoM$. Going even lower in $\lambda$, eventually (\ref{eq:throatstop})  will rise again, and the stop will cease to be tachyonic. This results in an excluded interval at intermediate $\lambda$, which we will call the ``throat" region. It exists as long as $\LoM$ is greater than some critical value:
\beq
\label{eq:throat}
 \lp\LoM\rp_{crit}  \sim \sqrt{\frac {3 g_3^2\lp 4 B_1 B_3 - B_2^2  \rp}{16 \pi^2 B_2 B_4} } 
\eeq
where we have approximated $h(\Lambda/M)\approx 1$.  

At small $\lambda$, the tuning is set by $\mu$ and $\Lambda_3$.  The tuning in the interesting regions of these models is everywhere a balance of $\mu$ and $\Lambda_\lambda$, and at higher values of $\lambda$ (i.e. on the horn region) the latter is leading.   The region of lowest tuning in these models sits roughly on the underside of the base of the horn.  This is sensible because one of the stops will run tachyonic near here, generating a larger $A_t/M_S$.  At the base, the tuning is set by $\mu$.   For two models (I.9 and I.13 -- both shown in fig.~\ref{fig:SquarkType}), a second region of comparably good tuning sits in the middle of the horn.  Although the stops are much heavier here, the tuning does not suffer greatly because the messenger scale has significantly decreased, and with it, $\abs{m_{H_u}^2}$ has decreased resulting in a lower value of $\mu$.  Other type I squark models exhibit some regions of parameter space with a similarly decreasing $\Lambda$, but the tuning is larger there.

We note that the least tuned type I squark models can achieve $\Delta_{FT}$ as low as $10^3$, and we stress that all of these models have regions of parameter space that are significantly less tuned than all type I Higgs and nearly all type II models.

\subsection{Type II couplings with mixing}

There are five type II (MSSM-MSSM-Messenger) couplings where the messenger mixes with one of the fields in an MSSM Yukawa coupling.  Three of these couplings are top-Yukawa-like, $Q U \phi_{5,H_u}$, $U H_u \phi_{10,Q}$ and $Q H_u \phi_{10,U}$, and two are bottom-Yukawa-like, $QH_d \phi_{\bar 5,D}$ and $QD\phi_{\bar 5, H_d}$.  The three top-Yukawa-like models are especially interesting because they provide an additional enhancement to $A_t$ by allowing two fields in  $\{ H_u, Q_3, U_3 \}$ to contribute to $A_t$, (\ref{eq:Atsum}).  This enhancement is so effective that one of these models, $UH_u \phi_{10,Q}$, is the least tuned of all models, possessing regions with $\Delta_{FT}\sim 850$. Overall, all three of the $y_t$ mixed models are significantly less tuned than the other type II models, because of this $A_t$ enhancement.

In order to present the formulas for any type II model (with a coupling of the form $\lambda X_1 X_2 \phi_{X_3}$) in a general simplified form, we first define $d_1\equiv d_{X_1}^{X_2X_3}$ (similarly for $d_2$, $d_3$).     We can derive an expression from (\ref{eq:typeIIsimp}) which can be applied to each type II model -- mixed or unmixed.\footnote{However, the $QQ\phi_{\bar D}$ can not be treated with these formulas due to the repeated $Q$.  As this model is one of the most tuned models, we will not address it in detail.  It is straightforward to derive the soft parameters in that model from (\ref{eq:typeIIsimp}).}    For these models, again with coupling $\lambda X_1 X_2 \phi_{X_3}$, we have,
\bea
\label{eq:TIIform}
\delta m_{X_1}^2 &=\frac{1}{256\pi^4} \lp d_1 \sum_i d_i\lambda^4 +2 d_1 d_3\lambda^2 y_{123}^2 - 2 d_1 C_r g_r^2 \lambda^2 
\right. \\ & -\left.   d_1^{2p} d_2 \lambda^2 y_{12p}^2 +\frac12 d_1 d_2^{pq} \lambda^2 y_{2pq}^2  \rp \Lambda^2 - \frac{d_1 \lambda^2 h\lp\LoM\rp}{96\pi^2 }\frac{\Lambda^4}{M^2} \\
\delta m_{X_2}^2 &=  \delta m_{X_1}^2 \{1\leftrightarrow 2 \} \\
\delta m_{X_a}^2 &= - \frac{1}{256\pi^4} \lp d_a^{1p}d_1y_{1ap}^2 + d_a^{2p}d_2y_{2ap}^2 \rp \lambda^2  \Lambda^2  \\
A_{X_{1,2}} &= - \frac{d_{1,2}}{16\pi^2} \lambda^2 \Lambda 
\eea
The $2 d_1 d_3\lambda^2 y_{123}^2$ piece of $\delta m_{X_1}^2$ vanishes unless there is MSSM-messenger mixing, i.e. both $\lambda X_1 X_2 \phi_{X_3}$ and $y_{123} X_1 X_2 X_3$ appear in the superpotential.  The $d_i$ values are tabulated for each type II coupling in table~\ref{table:TIImodels}.  We now turn our focus to the individual models.

\begin{table}[t]
\begin{center}
\begin{tabular}{|c|c|cccc|}
\hline
\#&Model & $d_1$ &$d_2$ & $d_3$  & $C_r$ \\ 
\hline \spacer
II.1 & $Q U \phi_{5,H_u}$  & 1 & 2 & 3 &   $\lp \frac{13}{30},\frac 32,\frac 83\rp$ \\ \spacer 
II.2 & $U H_u \phi_{10,Q}$  &  2 & 3 & 1 &   $\lp \frac{13}{30},\frac 32,\frac 83\rp$ \\ \spacer 
II.3 & $Q H_u \phi_{10,U}$  & 1 & 3 & 2 &   $\lp \frac{13}{30},\frac 32,\frac 83\rp$ \\ \spacer  
II.4 & $Q D \phi_{\bar5,H_d}$  & 1& 2 & 3 &   $\lp \frac{7}{30},\frac 32,\frac 83\rp$ \\ \spacer 
II.5 & $QH_d \phi_{\bar5,D}$   & 1& 3 & 2 &   $\lp \frac{7}{30},\frac 32,\frac 83\rp$ \\ 
\hline  \spacer  
II.6 & $Q Q \phi_{5,\bar D}$  & 2 & 2 & 4 & $\lp \frac{1}{10},\frac 32, 4\rp$   \\  \spacer  
II.7 & $U D \phi_{\bar5,D}$    &  2 & 2 & 2 & $\lp \frac{2}{5},0, 4\rp$ \\  \spacer  
II.8 & $Q L \phi_{\bar5,D}$   &1& 3 & 2 &   $\lp \frac{7}{30},\frac 32,\frac 83\rp$ \\  \spacer  
II.9 & $U E \phi_{5,\bar D}$ &1& 3 & 1 &   $\lp \frac{14}{15},0,\frac 83\rp$  \\  \spacer  
II.10 & $H_u D \phi_{24,X}$  & 3 & 2 & 1 &   $\lp \frac{19}{30},\frac 32,\frac 83\rp$   \\
\hline \spacer  
II.11 & $H_u L \phi_{1,S}$   & 1 & 1 & 2 & $\lp \frac{3}{10},\frac 32, 0\rp$  \\  \spacer  
II.12 & $H_u L \phi_{24,S}$  & 1 & 1 & 2 & $\lp \frac{3}{10},\frac 32, 0\rp$    \\ \spacer  
II.13 & $H_u L \phi_{24,W}$  & $\frac 32$ & $\frac 32$ & 1 & $\lp \frac{3}{10},\frac 72, 0\rp$  \\ \spacer  
II.14 & $H_uH_d \phi_{1,S}$   & 1 & 1 & 2 & $\lp \frac{3}{10},\frac 32, 0\rp$    \\  \spacer  
II.15 & $H_u H_d \phi_{24,S}$  & 1 & 1 & 2 & $\lp \frac{3}{10},\frac 32, 0\rp$   \\  \spacer  
II.16 & $H_u H_d \phi_{24,W}$   & $\frac 32$ & $\frac32$  & 1 & $\lp \frac{3}{10},\frac 72, 0\rp$  \\
\hline 
\end{tabular}
\caption{The definitions of the parameters appearing in the formulas for the type II models.  The first five entries are the type II models with MSSM-messenger mixing.  The $d_i$ values preserve the order of the fields in the model column, e.g. \ for model II.1 $d_1=d_Q^{U\phi_{H_u}}$,  $d_2=d_U^{Q\phi_{H_u}}$ and  $d_3=d_{H_u}^{QU}$.  As before, $C_r = c_r^{H_u}+c_r^{\phi_1}+c_r^{\phi_2}$ is the sum of the quadratic Casimirs.  Note that model II.6 can not be directly plugged into the formulas of (\ref{eq:TIIform}).}
\label{table:TIImodels}
\end{center}
\end{table}

{\it $QU\phi_{5,H_u}$} --- This model is unique in that it is not bounded at high $\lambda$ by any stop tachyons, although slepton tachyons provide a similar ceiling.   Both stops receive a substantial enhancement to their soft masses from the mixing generated $y_t^2\lambda^2$ term.  Unsurprisingly, this model occupies a parameter space with features very similar to many type I squark models.  However, unlike those models, it is least tuned at high $\lambda$ and high $\LoM$ just above the region where $A_t$ is large enough to contribute negatively to the $m_h$.  The tuning in this model is controlled by $\Lambda_\lambda$ and $\mu$ at high and low $\LoM$, respectively.  The tuning in this model is shown in fig.~\ref{fig:TypeII3and4}.

\begin{figure}[t]
\begin{center}
\includegraphics[scale=.69]{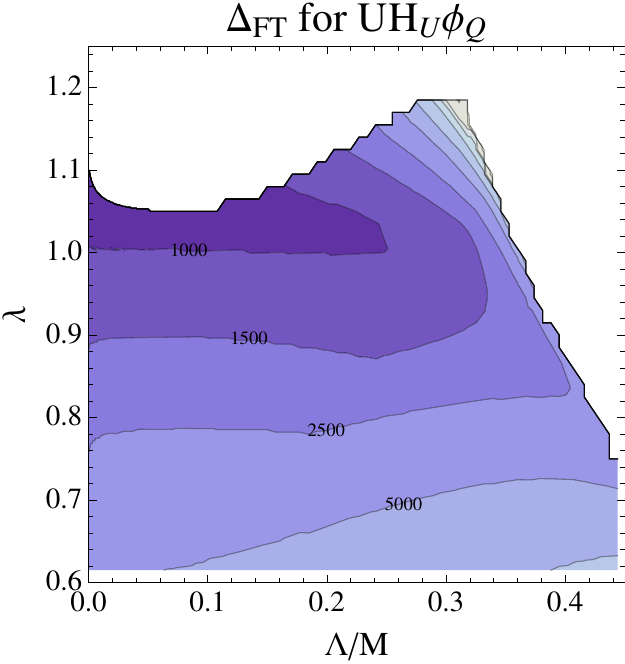}\qq
\includegraphics[scale=.71]{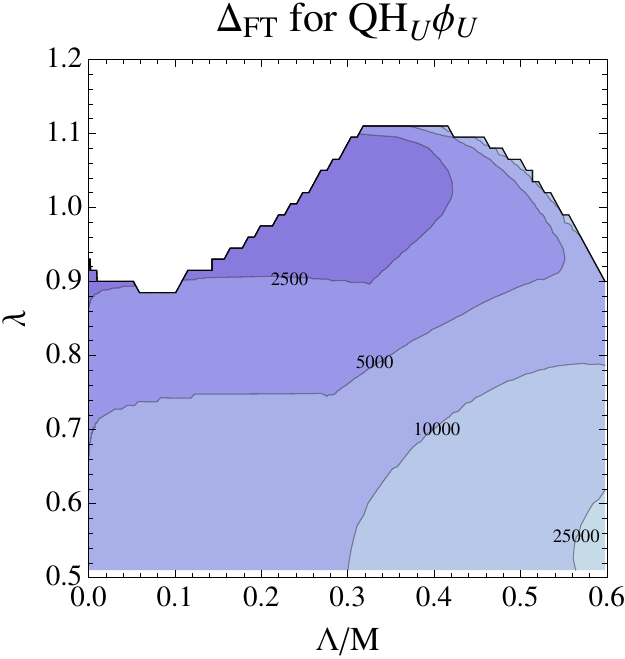}
\caption{The tuning shown for the two least tuned type II models, both of which have mixing between the messenger and an MSSM superfield. Left: $UH_u \phi_{10,Q}$. This model has mixing between the MSSM superfield, $Q$, and the messenger $\phi_{10,Q}$.  Right: $QH_u \phi_{10,U}$.  This model has mixing between the MSSM superfield, $U$, and the messenger $\phi_{10,U}$.  In both models, the tuning increases at high $\LoM$ because of the rising 1-loop contribution to $m_{H_u}^2$.  Both parameter regions are bounded on the right by stop tachyons. Above, they are also bounded by stop tachyons at higher $\LoM$, while at lower $\LoM$ the issue is unattainable EWSB due to the large positive contribution to $m_{H_u}^2$. 
}
\label{fig:TypeII}
\end{center}
\end{figure}

{\it $UH_u \phi_{10,Q}$} --- Due to contributions from both $A_U$ and $A_{H_u}$, this model receives the largest $A_t$ of any type II model, $A_t= \frac{5 \lambda^2}{16\pi^2}\Lambda$.  Additionally, this is the least tuned of any model -- type I or II.  The tuning contours are shown in fig.~\ref{fig:TypeII}.  This model possesses many aspects of  both the type I Higgs and squark models. $Q_3$ is tachyonic at a relatively low $\lambda$, however, the $U_3$ and $Q_3$ tachyons intersect near $\lp\lambda,\LoM\rp\sim\lp1.2,0.32\rp$, so no horn feature appears in this model.  As with the type I Higgs models, this model is least tuned for small $\LoM$ because the one loop contribution to $m_{H_u}^2$ is large everywhere else.  This leads a strip in $\LoM$ near $\lambda\sim1$ of lowest tuning.  This strip is cut off from above by the large positive contribution to $m_{H_u}^2$ causing problems with EWSB (as in type I Higgs models), grows more tuned to the right from the increasingly negative $m_{H_u}^2$ and more tuned below by an increasing messenger scale. In the region of lowest tuning, $\Lambda_3$ is the largest contributor.  For high $\LoM$ and high $\lambda$, tuning is controlled by the one-loop contribution.  At low $\lambda$, $\mu$ is the largest contributor to tuning.

{\it $QH_u \phi_{10,U}$} --- This model, shown in fig.~\ref{fig:TypeII}, is similar to $UH_u \phi_{10,Q}$, however the relation $d_U=2d_Q$ works against this model in two ways. First, the smaller multiplicity factor for $Q_3$ means that the $A$-term is not quite as large as in the $UH_u \phi_{10,Q}$ model.  Second, the larger multiplicity factor for $U_3$ gives a large contribution to $m_{H_u}^2$ from the mixing term $y_t^2\lambda^2$ and this contribution causes problems with EWSB at lower $\lambda$ compared to the $UH_u \phi_{10,Q}$ model.  In particular, this cuts into the region where $A_t/M_S\to \sqrt 6$, further reducing the quality of this model.  As in the $UH_u \phi_{10,Q}$ model, the tuning at low $\lambda$ is controlled by $\mu$.  At higher $\lambda$ values, $\Lambda_3$, $\Lambda_\lambda$ and $\Lambda_{1-loop}$ control the tuning for low, medium and high $\LoM$ respectively.

{\it $QH_d \phi_{\bar 5,D}$} --- This model lives in a short slice of parameter space bounded by $U_3$ tachyons above and below by regions where EWSB cannot be achieved. The tuning in the model is controlled by $\mu$ everywhere.   Since the mixing of this model gives additional enhancements proportional to $y_b^2$, the effect of the mixing leads to only a minuscule contribution which makes no appreciable change to the case where the term is absent.

{\it $QD\phi_{\bar 5, H_d}$} --- This model lives in a rather narrow slice of parameter space bounded by $U_3$ tachyons above and $D_3$ tachyons below.
The tuning in the model is governed everywhere by $\mu$ and $\Lambda_\lambda$.  As the in the previous model which mixes with the bottom Yukawa, no significant change manifests from the effect of the mixing.

\subsection{Type II couplings without mixing}

The type II couplings without mixing, which frequently have very small viable regions in parameter space, tend to have high tuning.  This is primarily because near a stop tachyon, $M_S$ is smaller (allowing for large $A_t/M_S$), but the viable regions in many type II models are sculpted in part by tachyons which are uncorrelated with $M_S$. Additionally, these models do not receive any significant $A_t$ enhancement, as was the case for models II.1-II.3 as discussed in the previous subsection. We now briefly discuss the features of each model.  

\begin{figure}[t]
\begin{center}
\includegraphics[scale=.68]{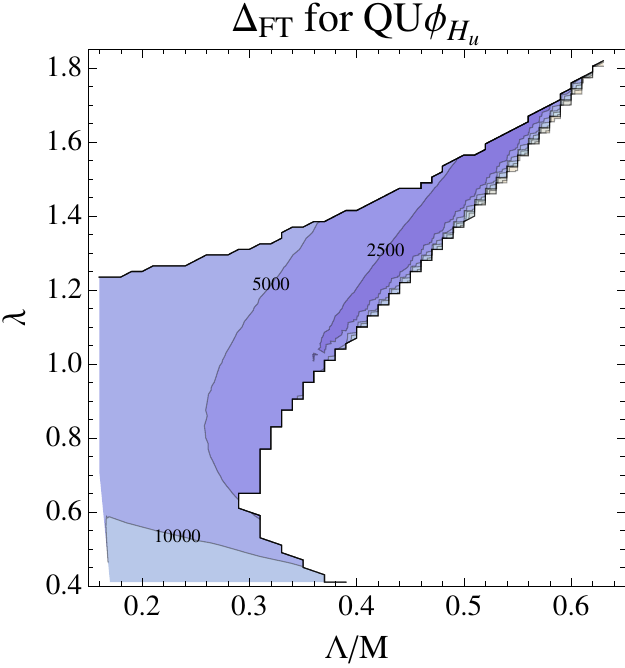}\qq
\includegraphics[scale=.72]{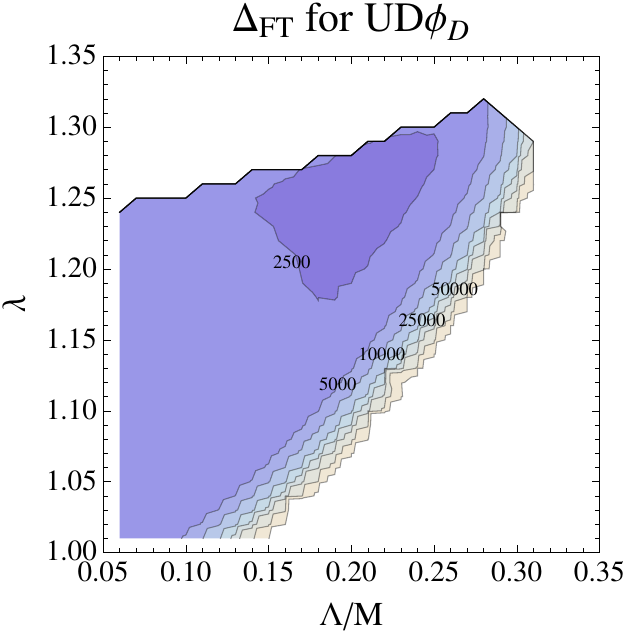}
\caption{The tuning shown for the third and fourth least tuned type II models,.  Left: $QU\phi_{\bar 5,\phi_{H_u}}$.  This model is the worst of the top-Yukawa-like mixed models.  Unsurprisingly, it also receives the smallest contribution to $A_t$.  It generally looks similar to the type I squark models in shape.  The tuning is best however at high $\LoM$    Right: $UD \phi_{\bar 5,D}$.  This model is bounded from above by tachyonic stops from $Q_3$.  The divergent region at the bottom of the plot is where $A_t/M_S$ is growing non-perturbative.  While much of the region has lower tuning, it exists in a very small window of $\lambda$ and $\LoM$ space. 
}
\label{fig:TypeII3and4}
\end{center}
\end{figure}

{\it $QQ\phi_{5,D}$} --- Naively, one might expect this model to significantly enhance $A_t$ and manifest with regions of low tuning.   However, it turns out that this model is prevented from ever achieving maximal mixing. First, since $Q_3$ contributes doubly to $m^2_{U_3}$,  this results in a maximum MSSM-messenger coupling of $\lambda\sim 0.8$ before $U_3$ tachyons enter.  Additionally, $Q_3$ itself is tachyonic for all $\LoM$ between $\lambda\sim\{0.45,1.1\}$.  This is due to the same contributions which induce a squark throat (i.e. the discriminant in (\ref{eq:throat}) is negative).  Thus, this model has no valid solutions in any region with $\lambda\gtrsim 0.45$.  The tuning in the small region that is valid in this model is governed by $\mu$ and $\Lambda_3$, however, the MSSM-messenger coupling contribution is truly negligible here.

{\it $UD\phi_{\bar 5, D}$} --- This model, with tuning shown in fig.~\ref{fig:TypeII3and4}, lives in a small slice of parameter space bounded by $Q_3$ tachyons above and $U_3/D_3$ tachyons below.  The regions with the best tuning in this model are at higher values of $\lambda$ near where $Q_3$ is tachyonic.  Encroaching on the region bounded below drives the SUSY breaking scale $\Lambda$ up rather drastically due to the $A_t/M_S$ growing so large that it provides a negative radiative correction to $m_h$ (this effect appears so pronounced in this model mostly due to the small parameter space considered).  Near the very large $\Lambda$ regions, the tuning is set by $\mu$, but in the regions of lower tuning, it is set by the $\Lambda_\lambda$ term.   Of all the unmixed type II models, this one presents with the lowest tuning. 

{\it $QL \phi_{\bar 5,D}$} --- Perhaps unsurprisingly, this model is very similar to the $QH_d \phi_5$ model in the previous subsection.   It lives in a short slice of parameter space bounded by $U_3$ tachyons above and by slepton tachyons below.  Here, the tuning is controlled by $\Lambda_\lambda$ and $\mu$.  

{\it $U E \phi_{5,\bar D}$} --- This model is narrow in $\lambda$, but stretches further into $\LoM$ than the other type II models before the $Q_3$ tachyons above, and the $\stau_R$ tachyons below close the region.   The tuning in the model is controlled by $\mu$ everywhere.

{\it $H_u D \phi_{24,X}$} --- This coupling involves $H_u$, $D$ and a superfield with gauge charges like the $X$ bosons of Pati-Salam models.  This model has similarities with both the Higgs and squark type I models.  At low $\LoM$, raising $\lambda$ causes issues with electroweak symmetry breaking.  To the right a slight $D_3$ tachyon throat cuts off the model.  The tuning is driven by $\mu$, $\Lambda_{1-loop}$ and $\Lambda_\lambda$.

{\it $H_u H_d \phi_{1,S},H_u H_d \phi_{24,S},H_u H_d \phi_{24,W}$} --- In addition to being horribly tuned, these models introduce a $\mu-B\mu$ problem.   Overall, the other features of these models is quite similar to the type I Higgs models.  Note that for $H_u H_d \phi_{1,S}$, the $\phi_1$ singlet field is insufficient to mediate SUSY breaking, so an additional $\mathbf{5}+\mathbf{\bar5}$ with no superpotential interactions with the MSSM is assumed.  The tuning is driven by $\Lambda_{1-loop}$ and $\mu$.

{\it $H_u L \phi_{1,S},H_u L \phi_{24,S},H_u L \phi_{24,W}$} --- These models have EWSB issues above and a slepton tachyon throat structure analogous to the squark throat discussed in section~\ref{sec:Squarks}.  However, more importantly these models introduce unacceptably large neutrino masses. 
Additionally, the $\phi_1$ singlet field is insufficient to mediate SUSY breaking, so an additional $\mathbf{5}+\mathbf{\bar5}$ with no superpotential interactions with the MSSM is assumed.  The tuning is driven by $\Lambda_{1-loop}$ and $\mu$.

\section{Phenomenology}
\label{sec:pheno}

In this section, we discuss aspects of the phenomenology of the models that give $\Delta_{FT}\lesssim 2 \times 10^3$.  The models that satisfy this are:  all type I squark models, the three mixed type II models --  $U_3H_u \phi_{10,Q}$,  $Q_3H_u \phi_{10,U}$ and $Q_3U_3 \phi_{5,H_u}$ -- and one other type II model -- $U_3D_3 \phi_{\bar 5,D}$.  Nearly all of these models possess spectra just beyond the reach of existing LHC searches.  It is interesting to note that such heavy spectra actually seem to be a {\it requirement} of these GMSB models with a 125 GeV Higgs.  This strongly suggests that the non-observation of SUSY and the presence of a heavy Higgs may be correlated issues rather than two distinct problems of SUSY.

\begin{figure}[t]
\begin{center}
\includegraphics[scale=1.83]{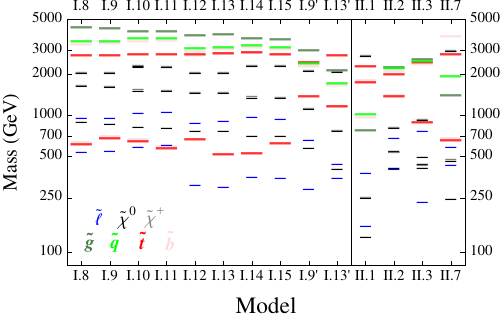}
\caption{The spectra for some of the better models at their points of least tuning are shown.  All type I squark models are shown to the left ($Q$: I.8-11 and $U$: I.12-15), type II models, including the three models which mix the top Yukawa with the messenger field and the $UD\phi_{D}$ (II.7) are shown to the right.  I.9$^\prime$ and I.13$^\prime$ denote the best point within the distinct region of comparable tuning accessible in these two models (see fig.~\ref{fig:SquarkType}) which present a very different spectra.  In the plot, thick, large lines denote colored particles -- $\go$, $\st_1$, $\st_2$, $\sbo_1$, $\sbo_2$ and $\tilde q$ (the nearly degenerate first-generation squarks) are shown.  The thinner lines denote uncolored particles -- $\tilde \ell$, $\cho^0$ and $\cho^\pm$ are shown.  All four neutralinos and both charginos are displayed.  In nearly all models, all right-handed sleptons and all left-handed sleptons/sneutrinos are approximately degenerate.
}
\label{fig:spectra}
\end{center}
\end{figure}

\subsection{Type I squark models}

In the region of least tuning (the base of the horn in fig.~\ref{fig:SquarkType}), the type I  squark  models have heavy gluinos and first generation squarks falling between $\sim$3.5-5 and $\sim$3-4.5 TeV respectively, while the lightest stop (as well as the sbottom in $Q_3$ models) has a mass between $\sim$0.5-1 TeV.  Additionally, there is almost always an NLSP $\stau$ or co-NLSP $\tilde \ell$s generally between $\sim$300-500 GeV  (although these sometimes appear even heavier than $700$ GeV).  However, the other region of low tuning appearing in models I.9 and I.13 (in the center of the horn) has a rather different profile (the best points of this second region are denoted by  I.9$^\prime$ and I.13$^\prime$ in fig.~\ref{fig:spectra}).  Here, the models have heavier stops, $\sim$1.2-2 TeV, but since $\Lambda$ has dropped significantly, the gluinos and first-generation squarks are now much lighter $\sim$2.0-3.5 TeV and $\sim$1.5-3 TeV, respectively.   Surveying these points with less tuning, it is clear that the mass of the lightest stop and the masses of the gluino and first-generation squarks tend to be anti-correlated -- thus, much of the parameter space with lower tuning possesses either squarks/gluinos or stops which can accessible at 14 TeV LHC.  

Given a stop portal, there are three separate simplified topologies that arise:
\begin{enumerate}
\item $\st$ ``NLSP'' --- with the stop appearing as either a co-NLSP with $\tilde \ell_R$, or as the true NLSP (e.g. the best point of  model I.11 in fig.~\ref{fig:spectra})
\item $\st$ : $\Bo$ : $\tilde \ell$ --- Decays of $\st \to t \cho^0 \to t \ell^\pm \tilde \ell^\mp$ 
\item $\st$ : $\tilde \ell$ --- This has competing decays of $\st \to b \nu \stau^+$ through an off-shell $\Ho$ and $\st \to  t \ell^\pm \tilde \ell^\mp $
\end{enumerate}
In all three cases, the NLSP decays to its SM partner plus gravitino. The first case is rare in these models, but has been discussed in the literature extensively, see e.g.~\cite{Chou:1999zb,Kats:2011it,Han:2012fw,Kaplan:2012gd,Kilic:2012kw}.  The second case, which is slightly more common, will populate high $\met$, high $H_T$ multilepton searches -- bounds for these will likely fall at or near the kinematic limit for $\st_1\st_1^*$ production.  The third case happens in models I.8-10,12-15 in fig.~\ref{fig:spectra}.  The third case is most interesting in the limit where the top decay is squeezed out ($m_\st-m_\stau <m_t$).  A sample spectrum which produces this signature is given by the best point of model I:14, as shown in fig.~\ref{fig:spectra}.

In this third case, when only the Higgsino-mediated decay can occur, the signature is two $b$-jets with moderate $p_T$, opposite sign $\tau$s with very high $p_T$ and large additional $\met$.  While in principle, several existing searches could have sensitivity to this exotic signature~\cite{ATLAS:2012ht,CMS-OS-DIL,CMS-PAS-EXO-12-002,MT2:2012uu}, those searches are not optimized to handle this kinematic configuration.  Existing opposite sign $\tau + \met$ searches tend to require a very hard jet and/or high $H_T$.  In the squeezed regime, the $b$-jets will generally lack substantial enough $p_T$ to meet these strict cuts.  Third generation leptoquark searches would have some sensitivity, however, the $b\tau$ invariant mass and $S_T$ requirements may be too harsh for this specific signature since so much of the event's energy is put into $\met$.  Searches for $M_{T_2}$ in the $t\bar t$ system are likely the most sensitive probe of this system.  Estimation based on the cuts in~\cite{MT2:2012uu} (as done for the similar $\{ b \tau^+ \nu\nu\} \{ \bar b\tau^- \nu\nu\}$ signal examined in~\cite{Evans:2012bf}) suggest this study repurposed would be sensitive to stops in the 450-550 GeV range.  A dedicated search region either for $\tau$s in the $M_{T_2}$ search or a hard $p_{T,\tau}$ region with softer cuts on the jets could likely probe significantly higher stop masses.   Further exploration of this signature is an exciting subject for future work.

The distinct regions of low tuning appearing with heavier stops and lower $\Lambda$ (in models I.9 and I.13) can have accessible gluinos and first-generation squarks (these spectra are denoted I.9$^\prime$ and I.13$^\prime$ in fig.~\ref{fig:spectra}).  These production portals result in complicated SUSY decay topologies with slepton co-NLSPs.  These models will generically give hard jets, large $\met$, and multiple leptons.  The 14 TeV extension of existing SUSY searches will be very sensitive to these topologies if the colored sparticles are light enough for a viable production cross-section.  Regions in other type I squark models exhibit the same decreasing $\Lambda$, but pay a substantial price in tuning.  While the spectra are not shown, they manifest with the same qualitative features -- rising stop masses with all other sparticles decreasing -- which can allow for first-generation squark/gluino production.

\subsection{Type II models}

As one can easily infer from fig.~\ref{fig:spectra}, the type II models have more variation in their signatures.  Generally these models possess lighter gluinos and first generation squarks.  In fact, the points of best tuning in model II.1, the mixed $Q_3U_3 \phi_{5,H_u}$ model, are already ruled out by existing searches because these colored particles appear near or even below 1 TeV.  However, in the other three less tuned models of type II, the gluinos are closer to 2 TeV and are not necessarily constrained by existing searches.  

The least tuned type II model, II.2 ($U_3H_u \phi_{10,Q}$), has stops and bottoms near 1.5 TeV, but the other squarks and gluinos are near 2 TeV.  For very low $\LoM$, these models will often have a $\Bo$ NLSP, however, the bulk of parameter space does have a right-handed slepton co-NLSPs.   In these low $\LoM$ regions, the SUSY breaking scale can be high enough to for a detector stable neutralino.  These models can present with classic SUSY signatures of multiple leptons, jets, $\met$ and very high $S_T$.  

The colored states of model II.3, $Q_3H_u \phi_{10,U}$, are slightly heavy, with stops near 1 TeV and both first-generation squarks and gluinos appearing near 2.5 TeV.  This spectra should be observable at 14 TeV LHC, where the production cross-section for 2.5 TeV gluinos and squarks is $\order{1 \mbox{ fb}^{-1}}$~\cite{Beenakker:2011fu}.  However, even moving into regions with slightly higher masses, these models still have very light slepton co-NLSPs appearing near 200-300 GeV, which could be easily observed at a future ILC.

A slightly different topology, with stops, gluinos and first-generation squarks near 750 GeV, 1.5 TeV and 2 TeV, respectively manifests in model II.7, $U_3D_3 \phi_{\bar 5,D}$.   Here, the Higgsino is very heavy, but the $\order{250\mbox{ GeV}}$ bino is always the NLSP.   As above, these complicated topologies cannot be seen until 14 TeV, but give rise to classic SUSY signatures of multiple leptons, jets, $\met$ and very high $S_T$.

\section{Conclusions}
\label{sec:conclusions}

In this work, we studied models that produce large $A$-terms through the introduction of a single marginal superpotential interaction between MSSM fields and the messengers of minimal GMSB. We classified all such interactions compatible with perturbative $SU(5)$ unification. Our complete list of 31 possible couplings -- 15 type I couplings (MSSM-messenger-messenger) and 16 type II couplings (MSSM-MSSM-messenger) -- is summarized in table \ref{table:models}.

Motivated by rampant confusion in the literature concerning the correct soft term contributions from MSSM-messenger interactions in the presence of MSSM-messenger mixing, we derived a new method for treating these interactions by directly integrating the wave functions in a manifestly holomorphic scheme. This conceptually straightforward method produced results applicable to all scenarios, whether or not fields are mixed.  Formulas were presented for the soft parameters both in complete generality (\ref{eq:Amsqfinal}); and in the simpler special cases of only type I or type II couplings, (\ref{eq:typeIgen})-(\ref{eq:typeI})
and (\ref{eq:typeIIgen})-(\ref{eq:typeIIsimp}) respectively.  

Using these new formulas and a slight variation on the Barbieri-Giudice tuning measure, we surveyed the tuning in each of the 31 models.  Under this examination, we concluded that the qualitatively similar type I Higgs models universally have high tuning due to the little $A/m_H$ problem, while the type I squark models can have tuning as low as $\Delta_{FT}\sim10^{3}$.   The majority of type II models have poor tuning, with the notable exception of the three models which allow for MSSM-messenger mixing with a top-Yukawa-like interaction, which generate a very large value for $A_t$.  The least tuned of these models, $U_3H_u\phi_{Q}$, manifests with the lowest tuning of any model studied in this work, $\Delta_{FT}\sim850$, while the other two still have tuning below $\Delta_{FT}<2\times 10^{3}$.

The spectra in the least tuned models usually have particles beyond the reach of the 8 TeV LHC, but can be accessible to the 14 TeV upgrade.  
This was not put in by hand, but is a consequence of the requirement that $m_h=125$ GeV, together with the minimal gauge mediated structure of the messenger sector. This suggests that the failure to find superpartners so far at the LHC was not an accident, but in fact had to be the case.

The models tend to possess either a stop below a TeV (and sometimes a sbottom as well) or accessible first-generation squarks and gluinos.  The decay chain in these scenarios terminates with the light NLSP stau,  co-NLSP sleptons or NLSP bino (with the rare exception of a stop NLSP) decaying to a gravitino. Both prompt and long-lived NLSP decays are possible in the models that we have considered. Production of the first-generation squarks and gluinos gives rise to classic SUSY signatures of high $\met$ and $S_T$ with multiple leptons and hard jets.  The scenarios with only stop production will usually given multi-lepton signatures; however, the cases with a stop NLSP or a squeezed $\st\to\stau^+ b \nu$ transition are more difficult.  The latter gives rise to a very interesting and poorly studied $b\bar b\tau^+\tau^- + \met$ signature. Uncovering search strategies to improve sensitivity to this final state is an exciting avenue for future study.  Another very common feature in these models is light sleptons (and occasionally Higginos) which could be readily produced and studied exhaustively at a TeV scale ILC.   

While we took an agnostic approach to flavor physics in this work (assuming a perfect alignment), the lack of MFV in all of the least tuned models begs a thorough treatment of flavor.   How much misalignment is permissible and whether alignment can be naturally achieved in a sensible way are both questions for future study. Another issue that we have not addressed in this paper is the origin of $\mu$ and $B_\mu$. It would be very interesting to extend the models considered here to include a mechanism for $\mu/B_\mu$. Perhaps extensions involving the NMSSM along the lines of \cite{CraigShih} are viable. Finally, while we assumed single MSSM-messenger interaction terms for simplicity in this paper, it would be interesting to explore the effects of including of multiple MSSM-messenger interactions, as this could potentially generate regions of parameter space which exhibit even lower tuning.

\section*{Acknowledgments}
\noindent
We thank  T.~Jeli\'nski, S.~Knapen, A.~Mariotti, Y.~Shirman, A.~Thalapillil, and B.~Zhu for useful conversations.  We thank B.~Allanach for help with SOFTSUSY.   JAE was supported by DOE grant DE-FG02-96ER40959. DS is supported in part by a DOE Early Career Award and a Sloan Foundation Fellowship. 

\appendix

\section{A Detailed Study of the $QU\Phi$ Model}
\label{sec:validation}

\subsection{Applying our general formulas}

In this appendix, we will provide an in-depth study of the model with
\begin{equation}
\label{eq:QUPhi}
W = \lambda Q U \Phi+y_t Q U H_u + X \Phi\tilde\Phi
\end{equation}
with $\langle X\rangle=M+\theta^2 F$. (We drop the 3rd generation subscript on $Q$ and $U$ to avoid cluttering the equations.) In this model, the messenger $\Phi$ has the same quantum numbers as $H_u$. This is in many ways the prime example of a mixed type II model, given that it has been studied already in many papers \cite{Shadmi:2011hs,Jelinski:2011xe,Abdullah:2012tq,Perez:2012mj,Evans:2011bea,Evans:2012hg,Endo:2012rd}. We will use this example to illustrate a number of points. First, it will highlight some interesting features of our general formulas. Second, by computing the soft masses in this model using other methods, it will provide a detailed check of our general formulas. Finally, we will use this example to illustrate the shortcomings of the formulas and approach in \cite{ChackoPonton} which were mentioned in the body of the paper.

As seen in (\ref{eq:typeIIsimp}) and (\ref{eq:TIIform}), the effect of MSSM-messenger mixing appears in the $\abs{\lambda}^2 \abs{y_t}^2$ terms, so to focus on that, let us for simplicity set the gauge couplings and all other MSSM Yukawa couplings to zero. We begin by quoting the result of the general formula (\ref{eq:TIIform}) for this model. Taking $X_1=Q$, $X_2=U$, $X_3=H_u$, and setting $g_r=y_b=y_\tau=0$, we have:
\bea
\label{eq:softmassesQUPhi}
A_{Q} &=-{1\over16\pi^2} d_{Q}  |\lambda|^2\Lambda\\
A_{U} &=-{1\over16\pi^2} d_{U}  |\lambda|^2\Lambda\\
\delta m_{Q}^2 &= {1\over256\pi^4}(d_Q (d_U +d_Q+d_{\Phi})|\lambda|^4 + 2d_Q d_{\Phi}  |\lambda|^2 |y_t|^{2})\Lambda^2\\
\delta m_{U}^2 &={1\over256\pi^4}(d_U(d_U+d_Q+d_{\Phi})|\lambda|^4 + 2d_U d_{\Phi} |\lambda|^2 |y_t|^{2})\Lambda^2\\
\delta m_{H_u}^2 &= -{1\over256\pi^4}d_{\Phi} (d_Q +d_U)|\lambda|^2|y_t|^2  \Lambda^2 
\eea
with no contribution to other soft masses. Here, as in section~\ref{sec:models}, $d_Q$ is shorthand for $d_Q^{U\Phi}$, etc. In the MSSM, we have $d_{Q}=1$, $d_{U}=2$ and $d_{\Phi}=3$, but it will be useful to leave these multiplicities general.
Notice that the last two $\abs{\lambda}^2\abs{y_t}^2$ terms from (\ref{eq:TIIform})  have cancelled out of $m_{Q}^2$ and $m_{U}^2$, leaving only the last line induced by the MSSM-messenger mixing. One can also check that by substituting the standard beta functions and anomalous dimensions into the formulas of \cite{ChackoPonton}, one misses these extra terms.   In the following subsections, we will study these extra terms in more detail. We will compute the soft mass-squareds in this model in two different ways: directly using SUSY correlators as in \cite{Craig:2013wga}, and directly using wavefunction renormalization in the interaction basis. The latter will also illustrate the subtleties of wavefunction renormalization which the method derived in section~\ref{sec:soft} avoids.

\subsection{Calculation using SUSY correlators}

Let us calculate the $m_i^2$ directly, using a supersymmetric correlator formalism along the lines of \cite{Craig:2013wga}. As in that paper, we separate out the $A$-term-squared contribution to $m_i^2$ coming from integrating out the auxiliary field $F_i$, and we will denote the remainder by the hatted quantity
\begin{equation}
m_i^2 = \hat m_i^2 + |A_i|^2
\end{equation}
This formulation is much simpler computationally, because it allows us to avoid various subtleties resulting from the treatment of contact terms and total derivatives.

The two-loop $\CO(|F|^2)$ contribution to $\hat m_Q^2$ is given by
\bea
\label{eq:mQZQ}
\hat m_Q^2 &=-|F|^2 \int d^4x_2\dots d^4x_6\,\Big\langle  \CQ^2(U(\lambda\Phi+y_t H_u))_1\bar\CQ^2(U^\dagger(\lambda^*\Phi^\dagger+y_t^* H_u^\dagger))_2  \\
& \qquad\qquad\qquad \CQ^2(Q U (\lambda\Phi+y_t H_u))_3 \bar\CQ^2(Q^\dagger U^\dagger (\lambda^*\Phi^\dagger+y_t^* H_u^\dagger))_4 (\Phi\tilde\Phi)_5 (\Phi^\dagger\tilde\Phi^\dagger)_6 \Big\rangle\\
&=-|F|^2\partial_M\partial_{M^*} \int d^4x_2\dots d^4x_4\,\Big\langle (U(\lambda\Phi+y_t H_u))_1 (U^\dagger(\lambda^*\Phi^\dagger+y_t^* H_u^\dagger))_2   \\
& \qquad\qquad\qquad \CQ^2( Q U (\lambda\Phi+y_t H_u))_3  \bar\CQ^2 ( Q^\dagger U^\dagger (\lambda^*\Phi^\dagger+y_t^* H_u^\dagger))_4  \Big\rangle \\
&\equiv -|F|^2\partial_M\partial_{M^*}  Z_Q^{(2)}(M,M^*)
\eea
These correlators are evaluated in the Euclidean supersymmetric free theory and only contain 1PI diagrams with respect to the theory containing the auxiliary fields $F_{Q,U,H_u}$ (in other words, if a diagram would become disconnected were an auxiliary field propagator removed, then it is not considered 1PI.) The subscripts $i=1,2,\dots$ are shorthand for the positions $x_i$. In the second equation, we have rotated the supercharges so that they act on the operators at $x_5$ and $x_6$, and then transformed this into $\partial_M\partial_{M^*}$ of a simpler correlator. Comparing with (\ref{eq:Amsq}), we can see that the correlator being differentiated is the 2-loop contribution to the wavefunction of $Q$. This is also clear diagrammatically, if we view this as the two-loop 1PI diagrams with external $F_QF_Q^\dagger$ in the presence of the interactions (\ref{eq:QUPhi}). 

It is straightforward to expand out all the terms in $Z_Q^{(2)}$ and perform the free-field contractions (keeping in mind the 1PI condition). The result is:
\bea
\label{eq:ZQI}
Z_Q^{(2)} &=   -|\lambda|^4 d_Q ( d_U I_1+d_\Phi I_2) - |\lambda|^2 |y_t|^2 d_Q( d_U(I_3+I_4)   + 2d_\Phi I_4)
\eea
where
\bea
& I_1 =  \int {1\over p^2(p^2+|M|^2)q^2((p+q)^2+|M|^2)},\quad & I_2 = \int   {1\over( p^2 +|M|^2)^2q^2(p+q)^2}\\
& I_3 = \int {1\over p^4 q^2((p+q)^2+|M|^2)},\quad & I_4 =\int {1\over p^2 (p^2+|M|^2) q^2(p+q)^2} 
\eea
with $\int$ shorthand for the integral over Euclidean phase space $ \int {d^4p\over (2\pi)^4}{d^4q\over (2\pi)^4}$. It is easy to check that
\begin{equation} 
\label{eq:Iders}
\partial_M\partial_{M^*} I_1 = \partial_M\partial_{M^*} I_2=-\partial_M\partial_{M^*} I_3 =\partial_M\partial_{M^*} I_4 = {1\over256\pi^4 |M|^2}
\end{equation}
We see that $I_3$ and $I_4$ contribute with equal magnitude and opposite sign to $\partial_M\partial_{M^*}Z_Q$. Thus the $|\lambda|^2|y_t|^2$ contribution proportional to $d_U$ drops out, consistent with what we found in (\ref{eq:softmassesQUPhi}) using the general formula. However, the contribution from mixing proportional to $d_\Phi$ remains. Combining  (\ref{eq:mQZQ}), (\ref{eq:ZQI}) and (\ref{eq:Iders}), and adding in $A_Q^2={d_Q^2\over (16\pi^2)^2}{|F|^2\over |M|^2}$, we find perfect agreement with (\ref{eq:softmassesQUPhi}).

Next, consider the soft mass for $U$. This can be found by the same manipulations used to derive $m_Q^2$, but with $d_Q\leftrightarrow d_U$. This agrees with the formula for $m_U^2$ in  (\ref{eq:softmassesQUPhi}).

Finally, we come to $m_{H_u}^2$. Here there is no $A$-term-squared contribution, and we have:
\bea
m_{H_u}^2 &=-|F|^2 \int d^4x_2\dots d^4x_6\,\Big\langle  \CQ^2(y_t Q U )_1\bar\CQ^2(y_t^* Q^\dagger U^\dagger)_2  
 \CQ^2(\lambda Q U \Phi)_3 \bar\CQ^2(\lambda^* Q^\dagger U^\dagger \Phi^\dagger)_4 (\Phi\tilde\Phi)_5 (\Phi^\dagger\tilde\Phi^\dagger)_6 \Big\rangle \\
 &=-|\lambda|^2 |y_t|^2 |F|^2\partial_M\partial_{M*} \int d^4x_2\dots d^4x_4\,\Big\langle  \CQ^2( Q U )_1\bar\CQ^2( Q^\dagger U^\dagger)_2  
 ( Q U \Phi)_3 ( Q^\dagger U^\dagger \Phi^\dagger)_4 \Big\rangle \\
 &=+ d_\Phi(d_Q+d_U)|\lambda|^2|y_t|^2 |F|^2\partial_M\partial_{M^*} I_3\\
 &=  -{d_\Phi(d_Q+d_U)\lambda^2y_t^2\over (16\pi^2)}{|F|^2\over |M|^2}
\eea
which again agrees perfectly with  (\ref{eq:softmassesQUPhi}).

\subsection{Analytic continuation method}

As another check of the general formulas, let us also perform the analytic continuation calculation by directly and explicitly integrating the wavefunctions, which can be easily done in this simple example. The trick is to do a unitary field redefinition to go to the interaction basis:
\begin{eqnarray}
\Phi_1 &&= {\lambda\over\sqrt{\lambda^2+y_t^2}} \Phi +{ y_t\over\sqrt{\lambda^2+y_t^2}} H_u \\
\Phi_2 &&=-{ y_t\over\sqrt{\lambda^2+y_t^2}}  \Phi +{\lambda\over\sqrt{\lambda^2+y_t^2}} H_u 
\end{eqnarray}
So we will study the equivalent theory defined at the scale $\Lambda$:
\begin{eqnarray}
&& W = \hat\lambda Q U \Phi_1  + \kappa_1 X \Phi_1\tilde\Phi + \kappa_2 X \Phi_2\tilde\Phi\\
&& K = Q^\dagger Q + U^\dagger U +\Phi_1^\dagger\Phi_1+\Phi_2^\dagger\Phi_2 +\tilde\Phi^\dagger\tilde\Phi
\end{eqnarray}
where 
\begin{equation}
\label{eq:hatlambdadef}
\hat\lambda=\sqrt{\lambda^2+y_t^2}\, , \quad \kappa_1 = {\lambda\over\hat\lambda}\, ,\quad  {\rm and}\,\,\,\, \kappa_2=-{y_t\over\hat\lambda}
\end{equation} 
In the interaction basis, $\Phi_1$ will receive wavefunction renormalization, but $\Phi_2$ will not.   This fact simplifies the calculation considerably.

Now we evolve this theory down to a lower scale.  As in the previous subsections, we again neglect gauge couplings and the other Yukawas for simplicity. The result is:
\bea
 W &= \hat\lambda Q U \Phi_1  + \kappa_1 X \Phi_1\tilde\Phi + \kappa_2 X \Phi_2\tilde\Phi\\
K &=Z_Q(t) Q^\dagger Q +Z_U(t) U^\dagger U +Z_{\Phi_1}(t) \Phi_1^\dagger\Phi_1+ \Phi_2^\dagger\Phi_2 +  \tilde\Phi^\dagger\tilde\Phi
\eea
where the wavefunctions obey the beta functions:
\begin{equation}
\label{eq:Zeq}
-{1\over2}{d\log Z_{i}(t)\over dt} = \gamma_{i}={1\over16\pi^2} d_i \hat\lambda(t)^2 
\end{equation}
for $i=Q,U,\Phi_1$. Here $\hat\lambda(t)=\hat\lambda Z_Q^{-1/2}Z_U^{-1/2}Z_\Phi^{-1/2}$ is the nonholomorphic Yukawa coupling in the theory with canonical K\"ahler potential. It obeys the beta function
\begin{equation}
\label{eq:lambdaeq}
{d\hat\lambda(t)\over dt}  =\hat\lambda(t)(\gamma_Q+\gamma_U+\gamma_\Phi)
\end{equation}
with boundary condition $\hat\lambda(0)=\hat\lambda$. These equations can be integrated to obtain $Z_i(t)$; the result is 
\begin{equation}
\log Z_i(t) = {d_i\over d_Q+d_U+d_\Phi} \log \left( 1-{(d_Q+d_U+d_\Phi)\hat\lambda^2\over 8\pi^2}t\right)
\end{equation}

Next, at the scale $M$, we  integrate out the messengers supersymmetrically. This sets $\tilde\Phi=0$ and $\Phi_2=-\kappa_1\Phi_1/\kappa_2$. We additionally identify $\Phi_1$ with $H_u$ below the messenger scale. At $M$ this theory then becomes
\begin{eqnarray}
&& W = \hat\lambda Q U H_u  \\
&& K = Z_Q(t_M)Q^\dagger Q + Z_U(t_M) U^\dagger U +\left(Z_{\Phi_1}(t_M)+{\kappa_1^2\over \kappa_2^2}\right)H_u^\dagger H_u
\end{eqnarray}
where $t_M\equiv \log |M|/\Lambda$.  It is clear that the wavefunction of $H_u$ is discontinuous at the messenger scale. This is precisely the issue alluded to in section~\ref{sec:soft} arising from the mixing of $H_u$ and $\Phi$ that prevents the formulas of \cite{ChackoPonton} from being applied as presented.

Finally, we should proceed with wavefunction renormalization by again integrating (\ref{eq:Zeq})-(\ref{eq:lambdaeq}), but now from $M$ to some lower scale $\mu$. The key difference is that because of the discontinuity in the wavefunction, the boundary condition for the Yukawa coupling is also discontinuous:
\begin{equation}
\hat\lambda(t_M^-)=\hat\lambda Z_Q(t_M)^{-1/2}Z_U(t_M)^{-1/2}\left(Z_{\Phi_1}(t_M)+{\kappa_1^2\over \kappa_2^2}\right)^{-1/2} \equiv \hat\lambda_M
\end{equation}
Taking this into account, we find
\bea
-\log Z_{Q,U}(\log\mu) &= -{d_{Q,U}\over d_Q+d_U+d_\Phi} \Bigg[ \log \left( 1-{(d_Q+d_U+d_\Phi)\hat\lambda^2\over 8\pi^2}\log {|M|\over\Lambda}\right) \\
&\qquad\qquad - \log \left( 1-{(d_Q+d_U+d_\Phi)\hat\lambda_M^2\over 8\pi^2}\log {\mu\over |M|}\right) \Bigg]\\
-\log Z_{H_u}(\log \mu) &=-\log\left( \left( 1-{(d_Q+d_U+d_\Phi)\hat\lambda^2\over 8\pi^2}\log{|M|\over\Lambda}\right)^{{d_{\Phi}\over d_Q+d_U+d_\Phi}} + {\kappa_1^2\over\kappa_2^2}\right)\\
&\qquad\qquad - {d_{\Phi}\over d_Q+d_U+d_\Phi}\log \left( 1-{(d_Q+d_U+d_\Phi)\hat\lambda_M^2\over 8\pi^2}\log {\mu\over |M|}\right)
\eea
Differentiating these expressions with respect to $M$ and $M^*$ and substituting (\ref{eq:hatlambdadef}) to recover the dependence on the original couplings, we again find perfect agreement with the general results   (\ref{eq:softmassesQUPhi}) for $m_{Q,U,H_u}^2$.

\section{Fine-Tuning Measure}
\label{sec:FT}

Fine-tuning is an inherently ambiguous concept.  When comparing variations of two nearly identical parameters, it seems sensible, however even when comparing the variation of a mass parameter to a coupling the choice of measure quickly looks somewhat arbitrary.   Since one of the objectives of this work is to determine which GMSB models possess lower fine-tuning, we demand that our fine-tuning measure, $\Delta_{FT}$, satisfy certain properties:
\begin{enumerate}
\item $\Delta_{FT}$ should provide a meaningful and accurate comparison between GMSB scenarios
\item $\Delta_{FT}$ should never overlook contributions from large terms which cancel in a uncorrelated way
\item $\Delta_{FT}$ should never introduce contributions from large terms which cancel in a correlated way
\item $\Delta_{FT}$ should assign comparable sensitivity to two uncorrelated terms which cancel one another
\end{enumerate}    
Traditionally, the Barbieri-Guidice tuning measure~\cite{GiudiceBarbieri} is defined as:
\beq
\Delta_{BG} \equiv \max \{ \Delta_a\} \;\; \mbox{ where }\;\; \Delta_a \equiv \frac{\d \log m_z^2}{\d \log a} 
\eeq
where $a$ sums over a set of ``fundamental" parameters. In general, it is not so clear what these fundamental parameters should be, and different choices for them lead to different numerical values of the tuning measure (the measure is not reparametrization invariant).  For our purposes, we will adapt the Barbieri-Giudice tuning measure to be,
\beq
\label{eq:FT}
\Delta_{FT} \equiv \max \{\Delta_i\} \;\; \mbox{ where } \;\; \Delta_i \equiv \frac{\d \log m_z^2}{\d \log \Lambda_i^2} 
\eeq
where $\Lambda_i \in \{ \lambda_i^2 \Lambda, \Lambda_{1-loop},\mu \}$ and  $\lambda_i$ runs over all the important couplings in the theory, i.e.\ in this case, $\lambda_i\in \{g_1,g_2,g_3,y_t,y_b,y_\tau,\lambda\}$, although in practice only variations in $g_3$, $y_t$ or $\lambda$ manifest deviations large enough to matter quantitatively for this study.  Thus, our set of parameters is $\Lambda_i \in \{ \Lambda_3, \Lambda_t, \Lambda_\lambda,  \Lambda_{1-loop}, \mu \}$.  As in~\cite{Feng:2012jfa}, we choose to differentiate $m_z^2$ in (\ref{eq:FT}) with respect to only parameters with mass squared units as this serves to better adhere to our requirement that canceling terms provide comparable sensitivities.   

With the exception of $\mu$ (which is calculated directly), we compute these derivatives by implementing a very small fractional change in the parameters injected at the messenger scale, run down to the low scale and measure the change in $m_z^2$ using SOFTSUSY.  $\Lambda_{1-loop}$, see (\ref{eq:1loop}), was chosen as the parameter to account for the dependence on $\LoM$ (i.e. we fractionally vary the term $\frac{\Lambda^2}{M}\sqrt{h\lp\LoM\rp}$ rather than varying $F$, $M$, $\LoM$ or some other combination of these parameters.).  To keep variations in $\Lambda_{1-loop}$ and $\Lambda_\lambda$ orthogonal, we keep the 1-loop term fixed when we vary $\Lambda_\lambda$.  This method removes the possibility for uncorrelated cancellations and correlated values being erroneously treated.  Additionally, it is well defined in all of the GMSB scenarios we present here and, in principle, can be easily translated to work for many other GMSB scenarios as well.


\bibliography{atermbib}
\bibliographystyle{JHEP}

\end{document}